\documentclass[12pt]{article}
\pdfoutput=1
\usepackage[utf8]{inputenc}      
\usepackage[T1]{fontenc}          
\usepackage{hyperref}
\usepackage{amsmath,amssymb,mathtools}
\usepackage{booktabs,tabularx}
\usepackage{graphicx,subfig}
\usepackage{xspace}
\usepackage[usenames]{xcolor}\definecolor{fscolor}{RGB}{44,118,255}
\usepackage{geometry}
\usepackage{todonotes}
\usepackage{listings}
\usepackage[absolute]{textpos}
\usepackage[many]{tcolorbox}
\usepackage{xparse}
\usepackage[font=small,labelfont=bf,format=plain,margin=0.05\textwidth]{caption}
\usepackage{bbm}
\usepackage{xspace}
\usepackage{tabularx}
\usepackage{collcell,booktabs}
\usepackage{multirow}
\usepackage{multicol}
\usepackage{soul}
\usepackage[capitalize]{cleveref}
\usepackage[separate-uncertainty,multi-part-units=single,list-units=single]{siunitx}

\allowdisplaybreaks

\evensidemargin \oddsidemargin
\newcommand{\cp}{{\ensuremath{\mathcal{CP}}}\xspace}
\newcommand{\fb}{\;\text{fb}\xspace}
\newcommand{\pb}{\;\text{pb}\xspace}
\newcommand{\gev}{\;\text{GeV}\xspace}
\newcommand{\tev}{\;\text{TeV}\xspace}

\newcommand{\jet}{\text{jet}\xspace}
\newcommand{\HiBo}{\texttt{HiggsBounds}\xspace}
\newcommand{\HiBov}[1]{\texttt{HiggsBounds-#1}\xspace}
\newcommand{\MA}{\texttt{MA5}\xspace}

\newcommand{\Code}[1]{\texttt{\detokenize{#1}}}
\newcommand{\dtypearray}[2]{{\color{gray}\texttt{#1}(#2)}}

\newcolumntype{P}[1]{>{\raggedleft\collectcell\Code}p{#1}<{\endcollectcell}}
\newcolumntype{R}{>{\collectcell\Code}r<{\endcollectcell}}

\usepackage[sorting=none,%
  citestyle=numeric-comp,%
  bibstyle=mynumeric,%
  giveninits=true]{biblatex}

\NewBibliographyString{refname}
\NewBibliographyString{refsname}
\DefineBibliographyStrings{english}{%
  refname = {Ref\adddot},
  refsname = {Refs\adddot}
}
\DeclareCiteCommand{\ccite}
  {%
  \ifnum\thecitetotal=1
    \bibstring{refname}%
  \else%
    \bibstring{refsname}%
  \fi%
  \addnbspace\bibopenbracket%
  \usebibmacro{cite:init}%
   \usebibmacro{prenote}}
  {\usebibmacro{citeindex}%
   \usebibmacro{cite:comp}}
  {}
  {\usebibmacro{cite:dump}%
   \usebibmacro{postnote}%
   \bibclosebracket}
\newrobustcmd*{\Ccite}{\bibsentence\ccite}

\begin{document}

\thispagestyle{empty}
\def\thefootnote{\fnsymbol{footnote}}

\begin{flushright}
DESY-21-143\\
LU TP 21-38
\end{flushright}
\vspace{3em}
\begin{center}
{\Large\bf Testing Exotic Scalars with \texttt{HiggsBounds}}
\\
\vspace{3em}
{
Henning Bahl$^a$\footnote{email: henning.bahl@desy.de},
Victor Martin Lozano$^a$\footnote{email: victor.lozano@desy.de},
Tim Stefaniak$^a$\footnote{email: tim.stefaniak@desy.de (former address)},
Jonas Wittbrodt$^b$\footnote{email: jonas.wittbrodt@thep.lu.se},
}\\[2em]
{\sl ${}^a$ Deutsches Elektronen-Synchrotron DESY, Notkestr. 85, 22607 Hamburg, Germany},\\
{\sl ${}^b$ Department of Astronomy and Theoretical Physics, Lund University, Sölvegatan~14A, 223~62~Lund, Sweden}.
\def\thefootnote{\arabic{footnote}}
\setcounter{page}{0}
\setcounter{footnote}{0}
\end{center}
\vspace{2ex}
\begin{abstract}
{}

The program \texttt{HiggsBounds} is a well-established tool for testing Beyond-the-Standard Model (BSM) theories with an extended Higgs sector against experimental limits from collider searches at LEP, Tevatron and LHC\@. Thus far, it could be applied to any neutral or charged Higgs bosons originating from the modified Higgs sector. Implicitly, these particles were assumed to exhibit a somewhat hierarchical Yukawa structure as present in the Standard Model, where in particular the couplings to first generation fermions could be neglected. In this work, we extend the \texttt{HiggsBounds} functionalities to go beyond these restrictions, thus making the code applicable to any neutral or charged BSM scalars. Moreover, we develop a new approach to implement experimental searches whose kinematic acceptance depends significantly on the values of the involved couplings. We achieve this by recasting the searches to general scalar models. Using this approach we incorporate relevant current experimental limits from LHC searches for exotic scalars, and present the implications of these limits for a dark matter scalar mediator model, a flipped Two-Higgs-Doublet Model and a supersymmetric model with $R$-parity violation.

\end{abstract}

\newpage
\tableofcontents
\newpage
\def\thefootnote{\arabic{footnote}}


\section{Introduction}%
\label{sec:intro}

The Higgs boson discovered at the LHC in 2012 is the first observed potentially elementary particle with spin-$0$, i.e., a \emph{scalar boson}. In the Standard Model (SM) of particle physics, this scalar boson arises from augmenting the theory with a scalar $\text{SU}{(2)}_L$ doublet field and associated renormalizable and gauge-invariant terms in the scalar potential. For a specific constellation of the scalar potential parameters the \emph{Higgs mechanism} leads to a spontaneous breaking of the electroweak (EW) symmetry and provides masses to the $W^\pm$ and $Z$ bosons. The detection of the Higgs boson is thus a crucial sign that this theory is at least approximately realized at the currently probed energy scales.

While the Higgs mechanism in the SM provides a plausible explanation for the broken EW gauge symmetry and the massiveness of its associated gauge bosons, both observational and theoretical puzzles --- for instance, the presence of dark matter (DM) in our Universe, the quantum description of gravity, and the naturalness (or hierarchy) problem --- remain unsolved in the SM framework, and thus motivate the postulation and experimental search for new physics beyond the SM (BSM).

Many of such BSM theories postulate the existence of additional scalar particles. A prime example is Supersymmetry (SUSY)~\cite{Nilles:1983ge,Haber:1984rc,Martin:1997ns} which introduces a new scalar boson for every SM fermion, leading to a plethora of new spin-0 states. Other models simply extend the SM scalar sector by additional scalar $\text{SU}{(2)}_L$ doublet or singlet fields, leading, for example, to the Two-Higgs-Doublet-Model (2HDM)~\cite{Gunion:1989we,Branco:2011iw}. Yet another approach are simplified BSM models, in which only a minimal particle content is introduced to address a particular problem. For instance, addressing the DM problem, one may only postulate the existence of a stable DM candidate particle and a mediator particle that interacts both with the DM and SM particles~\cite{Abdallah:2015ter}. This mediator could e.g.\ be a neutral gauge boson (``$Z'$ boson'') of a new gauge symmetry, or simply a new scalar particle.

The ATLAS and CMS experiments at the LHC follow various strategies to search for these additional scalar particles. In SUSY --- under the additional assumption of conserved $R$-parity or proton hexality~\cite{Dreiner:2005rd} --- the new supersymmetric particles can only be produced pairwise at the collider, and would then decay (possibly via cascades) into the lightest supersymmetric particle (LSP), which is electrically neutral and stable, thus escaping the detector unseen. SUSY searches, therefore, typically look for inclusive production of highly energetic objects (leptons, jets) associated with a fairly large amount of missing transverse energy (MET) (see \ccite{Schorner-Sadenius:2015cga,Canepa:2019hph,Zyla:2020zbs} for an overview of current searches). LHC searches targeting BSM models with an extended Higgs sector usually look for single production of a new scalar boson which either decays directly to SM particles ($t\bar{t}$, $b\bar{b}$, $\tau^+\tau^-$, $\mu^+\mu^-$, $W^+W^-$, $ZZ$, $\gamma\gamma$, etc.) or to final states containing also other scalars (often including the discovered Higgs boson). The third class of BSM models mentioned above --- the simplified DM models --- are often targeted by searches for a single highly-energetic object (jet, $Z$-boson, or Higgs boson, $h_{125}$) which recoils against the produced DM particles. The DM particles escape the detector and lead to MET~\cite{Kahlhoefer:2017dnp}.

Thus far, all these searches have not found any significant deviations from the SM background expectation. The results are, therefore, presented as upper limits on the signal cross section --- which rely on fairly modest model assumptions, e.g.\ the spin and \cp property of the involved new particle(s), specific coupling relations, or the decoupling of other BSM particles --- or as excluded parameter regions within specific BSM benchmark models. While the cross section limits can be applied directly to models that feature the same process with particles that fulfil the same assumptions, exclusion regions in BSM benchmark models can generally not be re-interpreted within different models. Therefore, if the model under study does not strictly fulfil all the assumptions underlying the presented search limit, a painstaking recasting analysis based on detailed Monte-Carlo (MC) simulations is often needed in order to estimate the sensitivity of the search and to derive a corresponding upper cross section limit on the alternative signal process. Naturally, such a recasting analysis cannot be as accurate as if the alternative signal process was directly analysed by the experimental collaboration. Ideally, the experimental collaborations would provide limits for all possible signal processes and parameter constellations their search is sensitive to, or, alternatively, provide additional information on how efficiencies and signal acceptances change under different model assumptions. Of course, resources and manpower of the experiments are limited, and the number of potentially relevant alternative signal processes can be large, rendering a complete coverage of all model-interpretations unrealistic in most cases.\footnote{Recommendations for the publication of experimental results have recently been put forward in \ccite{Abdallah:2020pec} as a guide to enable the maximal use of the experimental results.}

Several tools have been developed over the years to facilitate the assessment of the experimental validity of a new physics model, given the latest experimental limits from LHC searches. One such tool is \texttt{HiggsBounds}~\cite{Bechtle:2008jh,Bechtle:2010jt,Bechtle:2011sb,Bechtle:2013wla,Bechtle:2015pma,Bechtle:2020pkv} allowing to test extended Higgs sectors against upper limits from BSM Higgs searches. In this work, we present an extension of \HiBo\ that enables the program to test scalar particles that have quite different properties (to be quantified in detail later) than normally expected of BSM Higgs bosons. While for most BSM Higgs models the additional scalar particles couple only weakly to first and second generation quarks, this is not a necessity in generic scalar extensions of the SM\@. For specific high-energy models accommodating scalars with large first and second generation quark couplings see e.g.\ \ccite{Egana-Ugrinovic:2019dqu,Egana-Ugrinovic:2021uew}. Moreover, we test models involving quark-flavor and lepton-flavor violation.

First, this requires an extension of the \HiBo\ framework that handles the model predictions, in which hadronic cross sections are approximated from internal fit functions and effective couplings provided by the user. Second, new types of LHC search results had to be implemented. In particular, we considered limits from di-jet resonance searches~\cite{Aad:2019hjw,Aaboud:2018tqo,Aaboud:2018fzt,Sirunyan:2019vgj,Sirunyan:2018xlo,,Sirunyan:2018ikr,Khachatryan:2016ecr}, searches for di-jet resonances in association with an energetic photon~\cite{Aaboud:2019zxd} or lepton~\cite{Aad:2020kep}, and di-lepton resonance searches~\cite{Aad:2019fac,Aad:2020zxo,Aaboud:2018jff,Sirunyan:2018zhy,Sirunyan:2019shc}.

The ATLAS search for di-jet resonances in association with a photon from initial state radiation (ISR) is targeted to models with a $Z'$ boson. In order to make the limit applicable to scalar resonances we perform a detailed MC recasting analysis and derived cross section and signal acceptance functions for various constellations for the scalar-quark-quark couplings. All these results are incorporated in \HiBo\ in the form of simple fit formulae allowing the user to derive accurate and fast limits without running any MC generation himself. We describe our recasting analysis in detail in order to highlight important parameter dependencies and to motivate and guide the experimental collaborations to provide a corresponding signal interpretation in upcoming searches.

We demonstrate the extended \HiBo\ features and the impact of the newly implemented search limits in three exemplary model applications: first, a simplified DM model with a scalar mediator particle (``scalar DM portal model''); second, a 2HDM with large Higgs--$b$-quark couplings; third, we discuss the case of resonant scalar lepton (slepton) production and decay in SUSY models with $R$-parity violation.

This paper is organized as follows: In \cref{sec:simplified_model}, we discuss the generic scalar model we use for the implementation of all relevant di-jet and di-lepton searches. The implementation of these searches is then discussed in \cref{sec:searches} with a special focus on the di-jet plus photon search of \ccite{Aaboud:2019zxd} in \cref{sec:dijetISR}. In \cref{sec:results}, we discuss a simplified DM portal model, the 2HDM, and the $R$-parity violating MSSM as exemplary model applications.  The conclusions can be found in \cref{sec:conclusions}. \cref{sec:dijetISR_valid} provides details on the validation of our implementation of the ATLAS di-jet plus photon search. The derived fit formulas for the cross section and the acceptance can be found in \cref{sec:fits}. \cref{sec:HB} documents the new \HiBo routines implemented in the course of this work.


\section{Generic scalar models}%
\label{sec:simplified_model}

For the implementation of the di-jet and di-lepton limits, we employ generic scalar models. While these models are not meant to be complete BSM models, they are designed to allow for all relevant interactions.

For a neutral scalar $S$, we use the following generic scalar model,
\begin{equation}
\mathcal{L}_S =\begin{aligned}[t]& \frac{1}{2}\partial_\mu S \partial^\mu S - \frac{1}{2}m_S^2 S^2 \\
& - \frac{1}{\sqrt{2}}S \sum_{i,j=u,c,t}\left[\bar q_i (g_{q,ij} + i\gamma_5\tilde g_{q,ij})q_j +\text{h.c.}\right] \\
& - \frac{1}{\sqrt{2}}S \sum_{i,j=d,s,b}\left[\bar q_i (g_{q,ij} + i\gamma_5\tilde g_{q,ij})q_j +\text{h.c.}\right] \\
& - \frac{1}{\sqrt{2}}S \sum_{i,j=e,\mu,\tau}\left[\bar \ell_i (g_{\ell,ij} + i\gamma_5\tilde g_{\ell,ij})\ell_j  + \text{h.c.}\right],
\end{aligned} \label{eq:neutral_scalar_model}
\end{equation}
where the second and third line encode the couplings of $S$ to up-type and down-type quarks. The couplings $g_{q,ij}$ are the \cp-even Yukawa couplings to quarks; the couplings $\tilde g_{q,ij}$ are the \cp-odd Yukawa couplings to quarks. Note that we also allow for quark-flavor violating interactions. Similar to the quark couplings, we write down the couplings to leptons in the third line of \cref{eq:neutral_scalar_model} parameterized by the couplings $g_{\ell,ij}$ and $\tilde g_{\ell,ij}$ allowing for lepton-flavor violation.

The analogous generic scalar model for a charged scalar $S^\pm$ reads
\begin{equation}
\mathcal{L}_{S^\pm} =\begin{aligned}[t]& \partial_\mu S^\pm\partial^\mu S^\mp - m_{S^\pm}^2 S^\pm S^\mp \\
& - \frac{1}{\sqrt{2}} \sum_{i=u,c,t;j=d,s,b}\left[\bar q_i (g_{qL,ij} P_L + g_{qR,ij} P_R)q_j\cdot S^+ + \text{h.c.}\right]\\
& + \ldots \,,
\end{aligned}\label{eq:charged_scalar_model}
\end{equation}
where $P_{L,R}$ are the left- and right-handed chirality projection operators. The second line of \cref{eq:charged_scalar_model} encodes the interaction of the charged scalar with quarks allowing again for quark-flavor violation. The ellipsis in the last line of \cref{eq:charged_scalar_model} stands for interactions of $S^\pm$ with other SM particles (e.g.\ with a photon) or other BSM particles.

The general scalar model of \cref{eq:neutral_scalar_model,eq:charged_scalar_model} is available as \texttt{FeynRules} model file~\cite{Christensen:2008py,Degrande:2011ua,Alloul:2013bka}. This implementation is based on the \textit{DMsimp\_s\_spin1} \texttt{UFO} model~\cite{Abercrombie:2015wmb}. The model file is available as ancillary file accompanying the present paper.


\section{Implemented searches}%
\label{sec:searches}

In this Section, we discuss current experimental searches for BSM resonances decaying to di-lepton or di-jet final states. We implement all searches that either directly provided cross section limits or at least gave efficiencies for scalar resonances. Additionally, we use MC simulations in the generic models of \cref{sec:simplified_model} to obtain and tabulate the required efficiencies for scalar resonances in several models where they were not provided by the experiments.


\subsection{Di-lepton final states}

Higgs bosons decaying into $\tau^+\tau^-$ are among the most sensitive signatures in searches for models with extended Higgs sectors. As such, the existing searches in this channel, including the most recent CMS~\cite{Sirunyan:2018zut} and ATLAS~\cite{Aad:2020zxo} limits were previously implemented in \texttt{HiggsBounds}.\footnote{These $\tau\tau$ limits are implemented as exclusion likelihood profiles, see \ccite{Bechtle:2020pkv} for details.}

In contrast, decays into $e^+e^-$ and $\mu^+\mu^-$ pairs are strongly suppressed by their small masses for Higgs-like particles. While some searches involving $\mu^+\mu^-$ final states~\cite{ATLAS:2013qma, Aad:2014xva, Sirunyan:2019tkw, Aaboud:2019sgt, Sirunyan:2019bgz, Sirunyan:2017uvf} were already implemented in \HiBo, $e^+e^-$ signatures were not previously included at all. To allow implementing $e^+e^-$ searches in \HiBo, we extend the input framework to include this decay mode, see \cref{sec:HB} for details. This extension allows us to implement the latest ATLAS di-lepton resonance search~\cite{Aad:2019fac} which covers a huge, previously largely uncovered mass range from \SIrange{250}{3000}{\GeV} including finite width effects of up to \SI{10}{\%} of the resonance mass. The limit is set on a fiducial cross section and the required selection efficiencies for spin-0 resonances are available as auxiliary material in the analysis.

Limits on lepton flavor violation had previously only been implemented as limits on $\text{BR}(h_{125}\to e\mu/e\tau/\mu\tau)$~\cite{Aad:2019ugc, Aad:2019ojw}. We extended these by the latest lepton flavor violating resonance searches by ATLAS~\cite{Aaboud:2018jff} and CMS~\cite{Sirunyan:2018zhy,Sirunyan:2019shc} that all provide useable cross section limits for scalar particles.

For an overview of the newly implemented di-lepton searches see \cref{tab:dilepton_searches}.

\begin{table}\centering
  \begin{tabular}{lcccc}
    \toprule
    Channel                                           & Experiment & $\sqrt{s}$ [TeV] & Luminosity $[\text{fb}^{-1}]$ & Ref.               \\
    \midrule
    $pp \rightarrow X \rightarrow e^+e^-$             & ATLAS      & 13               & 139                           & \cite{Aad:2019fac}      \\
    $pp \rightarrow X \rightarrow \mu^+\mu^-$         & ATLAS      & 13               & 139                           & \cite{Aad:2019fac}      \\
    \midrule
    $pp \rightarrow X \rightarrow e\mu,e\tau,\mu\tau$ & ATLAS      & 13               & 36.1                          & \cite{Aaboud:2018jff}   \\
    $pp \rightarrow X \rightarrow e\mu $              & CMS        & 13               & 35.9                          & \cite{Sirunyan:2018zhy} \\
    $pp \rightarrow X \rightarrow \mu\tau,e\tau $     & CMS        & 13               & 35.9                          & \cite{Sirunyan:2019shc} \\
    \bottomrule
  \end{tabular}
  \caption{List of experimental searches for BSM resonances decaying into a di-lepton final states newly implemented into \HiBo.}%
  \label{tab:dilepton_searches}
\end{table}


\subsection{Di-jet final states}

Searches for di-jet resonances probe the highest masses accessible at the LHC\@. For scalars with a SM-Higgs-like coupling structure, they are typically not very sensitive, since at high masses the decays into top-quarks usually far outweigh those into light quarks and gluons. An exception are di-b-jet final states, which are commonly probed in searches for $b\bar{b}\to H \to b\bar{b}$~\cite{Aad:2019zwb, Sirunyan:2018taj}. This signature can be particularly sensitive to e.g.\ the high $\tan\beta$ regions of type II (e.g. MSSM) or flipped Yukawa sectors. However, di-$b$-jet searches are also carried out independently from dedicated Higgs searches, and we have implemented additional results from both ATLAS~\cite{Aaboud:2018tqo} and CMS~\cite{Sirunyan:2018ikr}. Furthermore, ATLAS performed a di-jet --- including di-$b$-jet --- resonance search for a resonance produced in association with a $\gamma$ from initial state radiation~\cite{Aaboud:2019zxd}. While the CMS di-$b$-jet result includes a useable interpretation for a scalar produced in either gluon fusion or $bb$-associated production, the two ATLAS results instead set limits on $\sigma\cdot A\cdot \epsilon$ for a Gaussian resonance. To implement these limits, the acceptances $A$ and efficiencies $\epsilon$ for a particular signal model need to be derived from Monte-Carlo simulations. In \cref{sec:dijetISR}, we discuss this procedure in detail for the di-jet + $\gamma$ search. The acceptances for the $b\bar{b}$ search~\cite{Aaboud:2018tqo} were obtained analogously to the more complicated case described in \cref{sec:dijetISR_acceptance_fit}.

Most searches for di-jet resonances that target gluon and light-quark jets --- with the current strongest in different mass ranges being~\cite{Khachatryan:2016ecr,Sirunyan:2018xlo,Aaboud:2018fzt,Sirunyan:2019vgj,Aad:2019hjw} --- only provide limits on a Gaussian resonance and no useable signal acceptance information for scalar signals. In principle, all of those searches could be implemented in \HiBo by deriving acceptances from Monte-Carlo simulations. However, this would be a considerable amount of work, in particular since most of those searches include finite-width effects. We, therefore, have not implemented those searches at this time, but strongly encourage the experimental collaborations to provide scalar acceptances --- in a similar way as derived in \cref{sec:dijetISR_acceptance_fit} --- for di-jet searches in the future.

Finally, ATLAS has also set a limit in the di-jet+$\ell$ final state~\cite{Aad:2020kep} that includes an interpretation as $pp \to t b (H^\pm \to tb)$. We have implemented this limit. It is, however, strictly weaker than the dedicated charged Higgs search~\cite{Aad:2021xzu} in the same channel and using the same dataset.

All newly implemented searches in di-jet final states are summarized in \cref{tab:dijet_searches}.

\begin{table}\centering
  \begin{tabular}{lcccc}
    \toprule
    Channel                                             & Experiment & $\sqrt{s}$ [TeV] & Luminosity $[\text{fb}^{-1}]$ & Ref.               \\
    \midrule
    $pp \rightarrow X \rightarrow \bar b b$             & ATLAS      & 13               & 36.1                          & \cite{Aaboud:2018tqo}   \\
    $pp \rightarrow X \rightarrow \bar b b$             & CMS        & 13               & 35.9                          & \cite{Sirunyan:2018ikr} \\
    $pp \rightarrow X + \gamma \rightarrow jj + \gamma$ & ATLAS      & 13               & 79.8                          & \cite{Aaboud:2019zxd}   \\
    $pp \rightarrow X \rightarrow jj + l$               & ATLAS      & 13               & 139                           & \cite{Aad:2020kep}      \\
    \bottomrule
  \end{tabular}
  \caption{List of experimental searches for BSM resonances decaying into a di-jet final states newly implemented into \HiBo.}%
  \label{tab:dijet_searches}
\end{table}


\subsection{ATLAS search for di-jet final state in association with initial-state photon radiation}%
\label{sec:dijetISR}

In this Section, we discuss the implementation of the ATLAS search for a BSM resonance decaying into a di-jet final state with initial-state photon radiation~\cite{Aaboud:2019zxd}. While the implementation of most other searches listed in \cref{tab:dilepton_searches,tab:dijet_searches} is comparably straightforward, the implementation of the di-jet plus photon search requires a dedicated effort.\footnote{The implementation of the ATLAS di-$b$-jet search~\cite{Aaboud:2018tqo} required a similar albeit simpler effort. No cross section fit was needed and the acceptance fit --- which was performed fully analogously to what is described in this section --- was considerably simplified as only one final state needed to be considered.} In \ccite{Aaboud:2019zxd} cross sections, acceptance, and efficiency values are only given for a simplified $Z'$ model, in which the $Z'$ couples to five lightest quarks with equal strength.

\begin{figure}
  \begin{minipage}{.48\textwidth}\centering
    \includegraphics[scale=1]{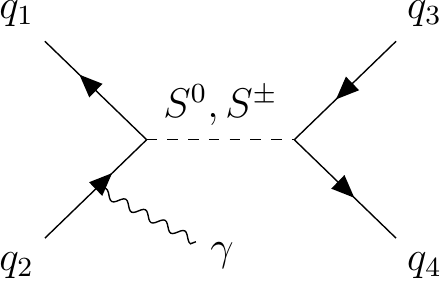}
  \end{minipage}
  \begin{minipage}{.48\textwidth}\centering
    \includegraphics[scale=1]{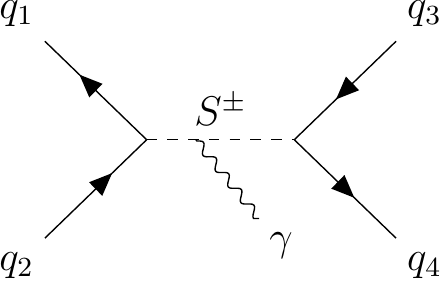}
  \end{minipage}
  \caption{Feynman diagrams for di-jet production via a scalar resonance in association with initial-state photon radiation.}%
  \label{fig:dijet_ISR_feynman}
\end{figure}

In order to make this experimental search usable within \HiBo (i.e.\ within the generic scalar model presented \cref{sec:simplified_model}, see also the corresponding Feynman diagrams in \cref{fig:dijet_ISR_feynman}), we need to derive the following dependencies of the cross section, the acceptance, and the efficiency:
\begin{itemize}
  \item Dependence on the different scalar-quark-quark couplings.

        While in \ccite{Aaboud:2019zxd} the resonance with universal couplings to all quarks (except the top quark\footnote{Due to the completely different kinematics, which we expect to lead to significantly lower acceptance values, decays into $t\bar{t}$ were not considered in the analysis. For the same reasons, we do not consider couplings involving top-quarks in our implementation as they would always lead to these unwanted decays in the given mass range. As a consequence, we also neglect production through gluon fusion as this would require significant top-quark couplings to be relevant. This approach is conservative in the sense that taking the additional decay and production channels into account would only enhance the sensitivity of the implemented analysis.}) was considered, we want to allow for a non-universal coupling structure and also take into account quark-flavor-violating couplings.

  \item Dependence on the electric charge of the resonance.

        While in \ccite{Aaboud:2019zxd} the considered resonance was assumed to have no electric charge, we want to consider also the case of an electrically charged resonance allowing for the radiation of the photon from the resonance itself (see right diagram of \cref{fig:dijet_ISR_feynman}).

  \item Dependence on the spin character.

        While in \ccite{Aaboud:2019zxd} only the example of a spin-1 resonance is discussed, we want to implement the case of a spin-0 resonance.

  \item Dependence on the \cp character of the decaying resonance.

        While the spin-1 resonance considered in \ccite{Aaboud:2019zxd} was considered to have only \cp-odd couplings, we want to allow for \cp-even and \cp-odd couplings of our spin-0 resonance.
\end{itemize}
In principle, a separate Monte Carlo simulation and a subsequent event analysis, reproducing the experimental analysis, has to be run for every parameter point and every possible quantum numbers of the resonance. This is not feasible for large parameter scans as they are often performed with \HiBo. While for the detector efficiency, we employ the values given in the auxiliary material of \ccite{Aaboud:2019zxd},\footnote{Here, we follow a recommendation from members of the ATLAS collaboration, which performed the analysis of \ccite{Aaboud:2019zxd}.}, more work is needed to work out fit functions for the cross section and the acceptance encoding the dependencies on all relevant parameters.


\subsubsection{Fitting the cross section}%
\label{sec:dijetISR_XS_fit}

For evaluating the cross section for di-jet production in association with a photon, we employ \texttt{MadGraph5\_aMC@NLO~2.7.0}~\cite{Alwall:2014hca} using the \texttt{NNPDF3.0 LO} PDF set~\cite{Ball:2012cx}. As model file, we employ the \texttt{UFO} version~\cite{Degrande:2011ua} of the models presented in \cref{sec:simplified_model}, which we generated using \texttt{FeynRules}.

Using this setup, we obtain the cross section for scalar production in association with a photon at leading oder (LO)\footnote{In principle, a NLO QCD calculation of the cross section would be possible with \texttt{MadGraph5} by extending the UFO model. We only consider the LO cross section, however, both for simplicity and for consistency with the acceptances that we want to validate against the LO results given in \ccite{Aaboud:2019zxd}.} as a function of the resonance mass and its couplings to quarks. We calculate this cross section for multiple parameter points deriving simple fit formulas (listed in \cref{sec:fits}) which we then implement into \HiBo. Within \HiBo, this cross section is then multiplied by the branching ratio of the scalar resonance to a di-jet final state.

\begin{figure}\centering
  \begin{minipage}{.48\textwidth}\centering
    \includegraphics[width=\textwidth]{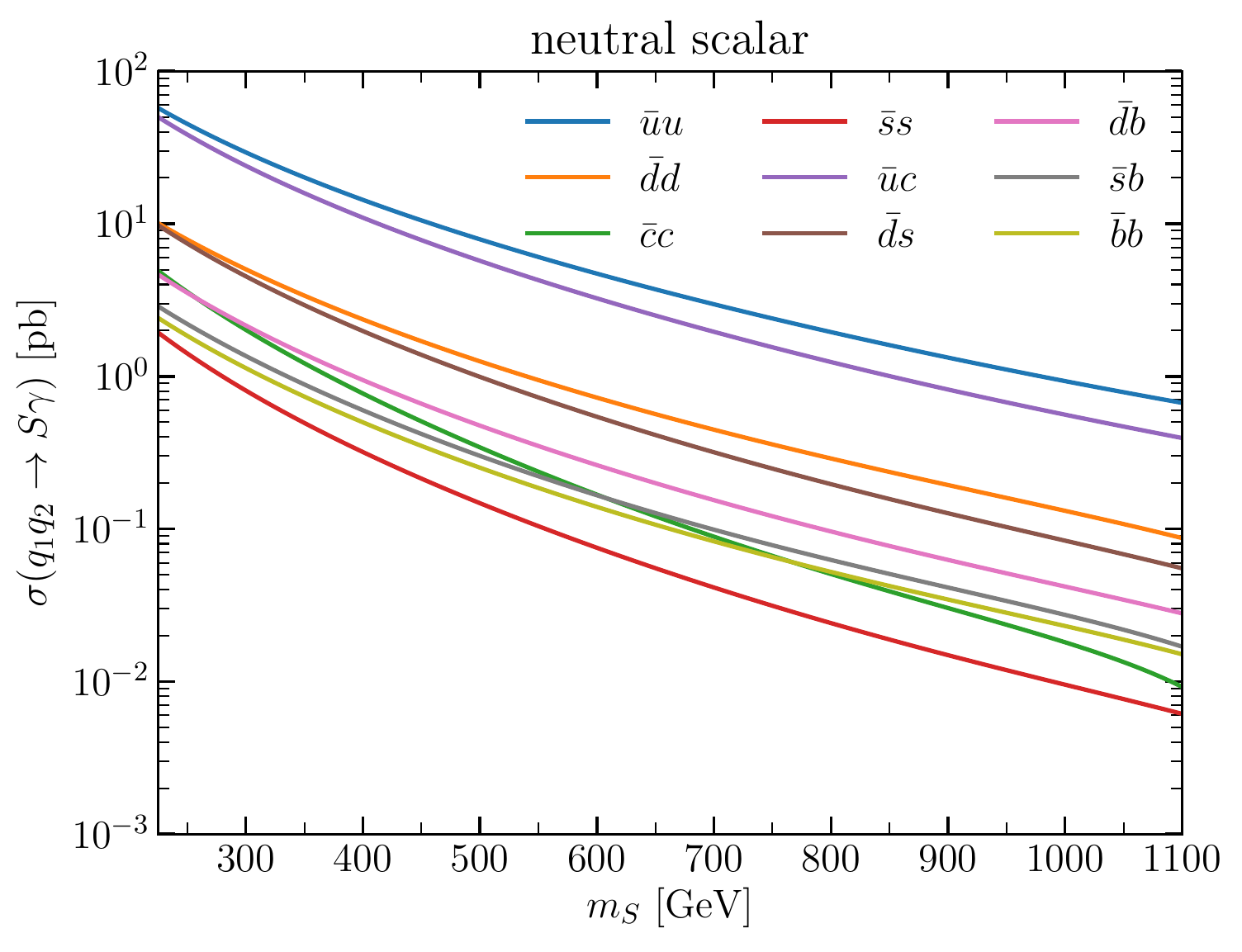}
  \end{minipage}
  \begin{minipage}{.48\textwidth}\centering
    \includegraphics[width=\textwidth]{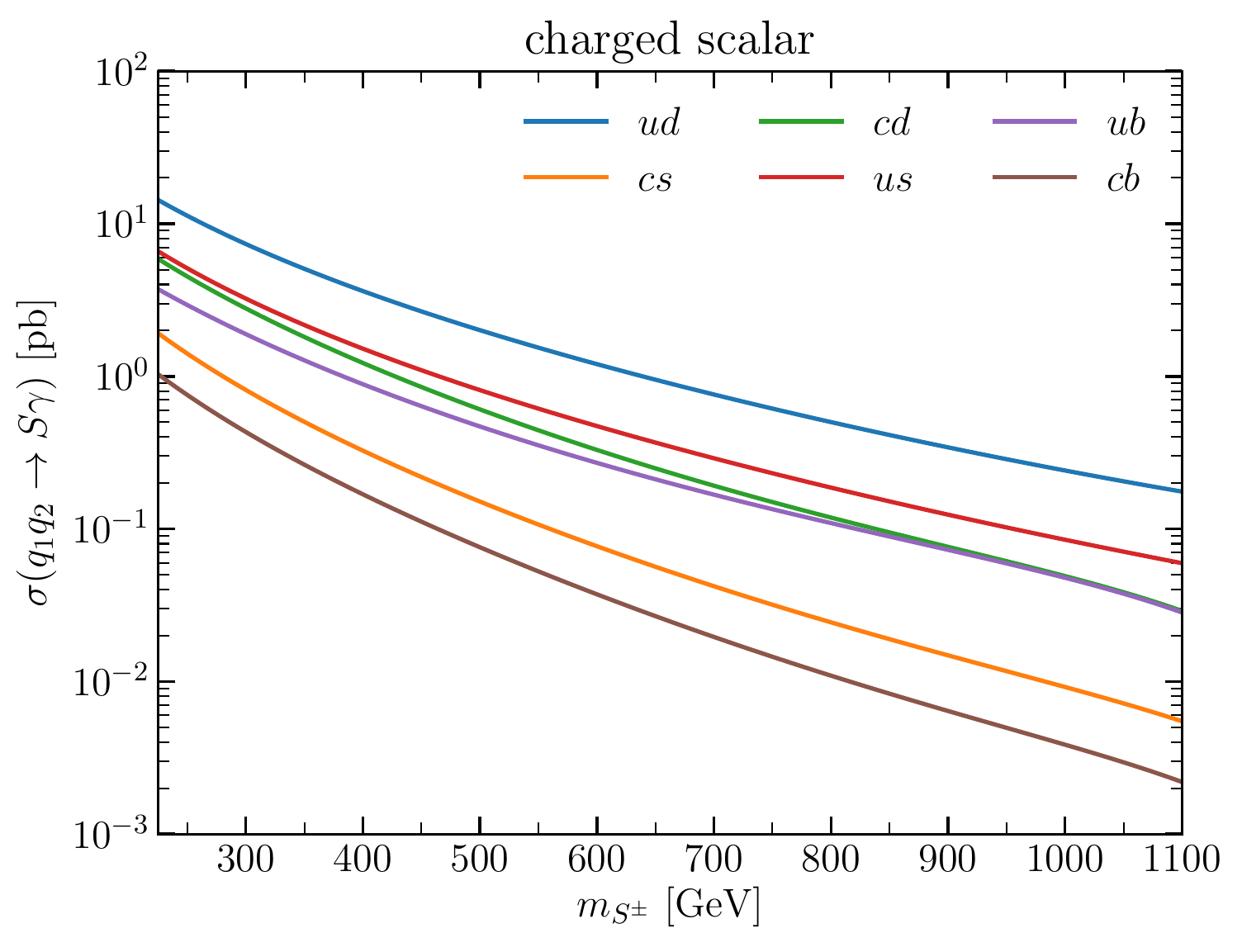}
  \end{minipage}
  \caption{Cross section of quark-initiated scalar production in association with a photon. \textit{Left:} neutral scalar. \textit{Right:} charged scalar (summing the positive and negative charged scalar cross sections).}%
  \label{fig:dijet_ISR_XS}
\end{figure}

In \cref{fig:dijet_ISR_XS}, we show exemplary cross-section results.\footnote{Here, the photons are required to have a transverse momentum of at least 10~GeV in order to avoid collinear singularities.} In the left panel, the cross-section for quark-initiated neutral scalar production in association with a photon is shown in dependence of the resonance mass. The scalar particle is assumed to have only one non-zero coupling to quarks. The non-zero coupling, which we set to one, is indicated in the plot legend. The cross section is highest if the quarks to which the resonance couples are abundant in the initial-state protons --- i.e., if no PDF suppression occurs --- and if up-type quarks, which have a higher electrical charge, are involved. Consequently, the cross section is highest if the resonance couples to up quarks (blue) reaching values of up to $\sim 60\pb$ for $M_S = 225\gev$, closely followed by the cross section for a resonance coupled to up and charm quarks (purple). All other cross sections are significantly lower. E.g., the cross section for a resonance coupled to down quarks (orange) is reduced by a factor of $\sim 5$ in comparison to the up-quark-induced and up-charm-quark-induced processes.

The behavior is quite similar for the production of a charged scalar in association with a photon (see right panel of \cref{fig:dijet_ISR_XS}). The overall size of the cross sections is comparable to the neutral scalar cross sections. The up-down-quark induced channel has the highest cross section reaching values of up to $\sim 10.5\pb$ for $M_S = 225\gev$.


\subsubsection{Fitting the acceptance}%
\label{sec:dijetISR_acceptance_fit}

To derive the analysis acceptance, we implement the analysis cuts of \ccite{Aaboud:2019zxd} into the \texttt{MadAnalysis5} (\MA) framework~\cite{Dumont:2014tja,Conte:2012fm,Conte:2014zja,Conte:2018vmg}.\footnote{The analysis code is publicly available as part of the \MA public analysis database~\cite{atlas_exot_2018_05}.} For the event generation, we employ the same setup as in \cref{sec:dijetISR_XS_fit}. To emulate the detector effects, we smear the invariant mass distribution of the final-state jets by the detector mass resolution as given in the auxiliary material of \ccite{Aaboud:2019zxd}.

\begin{table}\centering
  \begin{tabular}{l|cc}
    Criterion               & Single-photon trigger                              & Combined trigger      \\ \hline
    Number of jets          & \multicolumn{2}{c}{$n_\text{jets} \ge 2$}                                  \\
    Number of photons       & \multicolumn{2}{c}{$n_\gamma \ge 2$}                                       \\
    Leading photon          & $E_T^\gamma > 150\gev$                             & $E_T^\gamma > 95\gev$ \\
    Leading, subleading jet & $p_T^\jet > 25\gev$                                & $p_T^\jet > 65\gev$   \\
    Jet centrality          & \multicolumn{2}{c}{$|y^*| = |y_1 - y_2|/2 < 0.75$}                         \\
    Invariant jet mass      & $m_{jj} > 169\gev$                                 & $m_{jj} > 335\gev$
  \end{tabular}
  \caption{Cuts applied after trigger selection.}%
  \label{tab:dijet+photon_cuts1}
\end{table}

\begin{table}[t]
  \centering
  \begin{tabular}{l|cc}
    Criterion         & Inclusive           & $b$-tagged               \\ \hline
    Jet $|\eta^\jet|$ & $|\eta^\jet| < 2.8$ & $|\eta^\jet| < 2.5$      \\
    $b$-tagging       & --                  & $n_{b\text{-tag}} \ge 2$
  \end{tabular}
  \caption{Cuts applied to distinguish \textit{inclusive} and \textit{$b$-tagged} signal regions.}%
  \label{tab:dijet+photon_cuts2}
\end{table}

The analysis has been performed using two distinct data sets recorded using two different triggers:
\begin{itemize}
  \item a single-photon trigger requiring at least one photon candidate with $E_{T,\text{trig}}^\gamma > 140\gev$, and
  \item a combined trigger requiring at least one photon candidate with $E_{T,\text{trig}}^\gamma > 85\gev$ and two jet candidates, each with $p_T > 50\gev$.
\end{itemize}
The single-photon trigger is used for signal masses from 225~GeV to 450~GeV\@; the combined trigger from 450~GeV to 1100~GeV.

The photons and jets used within the actual analysis are then defined using stricter requirements. The photons are required to fulfill
\begin{itemize}
  \item $|\eta^\gamma| < 2.38$ excluding $1.37 < |\eta^\gamma| < 1.52$,
  \item $E_T^\gamma > 95 \gev$, and
  \item \textit{tight identification} and \textit{tight isolation} criteria~\cite{Aaboud:2018yqu,Aaboud:2018ugz,Aaboud:2017yyg}.
\end{itemize}
Jets are reconstructed using the anti-$k_t$ algorithm~\cite{Cacciari:2008gp} with radius parameter $R = 0.4$. After reconstruction, jets are required to fulfill
\begin{itemize}
  \item $|\eta^\jet| < 2.8$,
  \item $p_T^\jet > 25 \gev$, and
  \item a jet within $\Delta R = 0.4$ of a photon is removed.
\end{itemize}
Based upon these objects a series of selection cuts is applied to the events passing the trigger selection. These are summarized in \cref{tab:dijet+photon_cuts1}. In addition, the selections based on the single-photon and the combined trigger are further subdivided into an inclusive and a $b$-tagged signal region by applying the cuts displayed in \cref{tab:dijet+photon_cuts2}.

We validated our analysis by comparing to the example numbers given in \ccite{Aaboud:2019zxd} for the simplified $Z'$ model. More details can be found in \cref{sec:dijetISR_valid}.

For our present paper, it is, however, more appropriate to employ the limits for a generic Gaussian resonance given in \ccite{Aaboud:2019zxd} instead of the limits given for the simplified $Z'$ model. Correspondingly, we follow the procedure outlined in App.~A of \ccite{Aad:2014aqa}: Instead of retaining the full tail of the $m_{jj}$ distribution (see \cref{tab:dijet+photon_cuts1}), all events with a $m_{jj}$ value between $0.8 M$ and $1.2 M$ are selected where $M$ is the signal mass. The mean mass for the events within this interval is then used to read off the limit from the results given in \ccite{Aaboud:2019zxd}.

Using this analysis setup, we can derive the acceptance as a function of the resonance mass and the resonance's couplings to quarks. Since the simplified models as defined in \cref{eq:neutral_scalar_model,eq:charged_scalar_model} involve a relatively high number of free parameters, it is quite costly to run the MC generation and analysis chain for a sufficient number of parameter points in order to derive a fit formula for the acceptance in dependence of all relevant parameters. To avoid this, we first checked whether the acceptance depends on the \cp character of a neutral scalar resonance, or in case of a charged resonance on the chirality of its couplings to quarks. We find that these dependencies can be neglected (finding variations of $\lesssim 0.1\%$).

As a further simplification, we make use of the following observation: For almost all analysis cuts, the dependence on the flavors of the final-state quarks can be neglected. Therefore, the cut efficiencies can be approximated by a function depending only on the initial-state quark flavors. The only exceptions are the $m_{jj}$ cut as well as requirement of two $b$-tagged jets in the case of the $b$-tagged signal region. These cuts can be approximated by a function depending only on the flavors of the final-state quarks while the dependence on the flavors of the initial-state quarks can be neglected.

This approximate factorization can be written down in the form
\begin{align}
  \mathcal{A}_\text{incl-SR}(m_S,q_1,q_2,q_3,q_4) & \approx \mathcal{A}_\text{initial}(m_S,q_1,q_2) \cdot \mathcal{A}_{m_{jj}}(M_S,q_3,q_4), \label{eq:Aincl_factorization}              \\
  \mathcal{A}_\text{$b$-SR}(m_S,q_1,q_2,q_3,q_4)  & \approx \mathcal{A}_\text{incl-SR}(m_S,q_1,q_2,q_3,q_4) \cdot \mathcal{A}_{b\text{-tag}}(M_S,q_3,q_4),\label{eq:Abtag_factorization}
\end{align}
where $\mathcal{A}_\text{incl-SR}$ is the acceptance of the inclusive signal region and $\mathcal{A}_\text{$b$-SR}$ the acceptance of the $b$-tagged signal region. $\mathcal{A}_\text{initial}$ is the cut efficiency of the initial cuts (excluding the $m_{jj}$ and the $b$-tagging cuts); $\mathcal{A}_{m_{jj}}$, the cut efficiency of the $m_{jj}$ cut; and, $\mathcal{A}_{b\text{-tag}}$ is the cut efficiency of the $b$-tagging cut. $q_1$ and $q_2$ are the initial-state quarks; $q_3$ and $q_4$, the final-state quarks (see also \cref{fig:dijet_ISR_feynman}).

We cross-checked the approximation of \cref{eq:Aincl_factorization,eq:Abtag_factorization} for several example points (including cases with different initial and final state quark flavours) by directly evaluating the values for $\mathcal{A}_\text{incl-SR}$ and $\mathcal{A}_\text{$b$-SR}$ finding absolute deviations of $\lesssim 0.3\%$. We do not expect larger deviations for other parameter points.

The functions $\mathcal{A}_\text{incl-SR}$, $\mathcal{A}_{m_{jj}}$, and $\mathcal{A}_{b\text{-tag}}$ are then fitted by performing MC event generation and analysis for parameter points with different mass and coupling values. Since each of the functions depends only on two quark flavors --- or equivalently on one scalar-quark-quark coupling (see \cref{eq:neutral_scalar_model,eq:charged_scalar_model}) --- far less parameter points have to evaluated in comparison to fitting the functions $\mathcal{A}_\text{incl-SR}$ and $\mathcal{A}_\text{$b$-SR}$ directly.

\begin{figure}\centering
  \begin{minipage}{.48\textwidth}\centering
    \includegraphics[width=\textwidth]{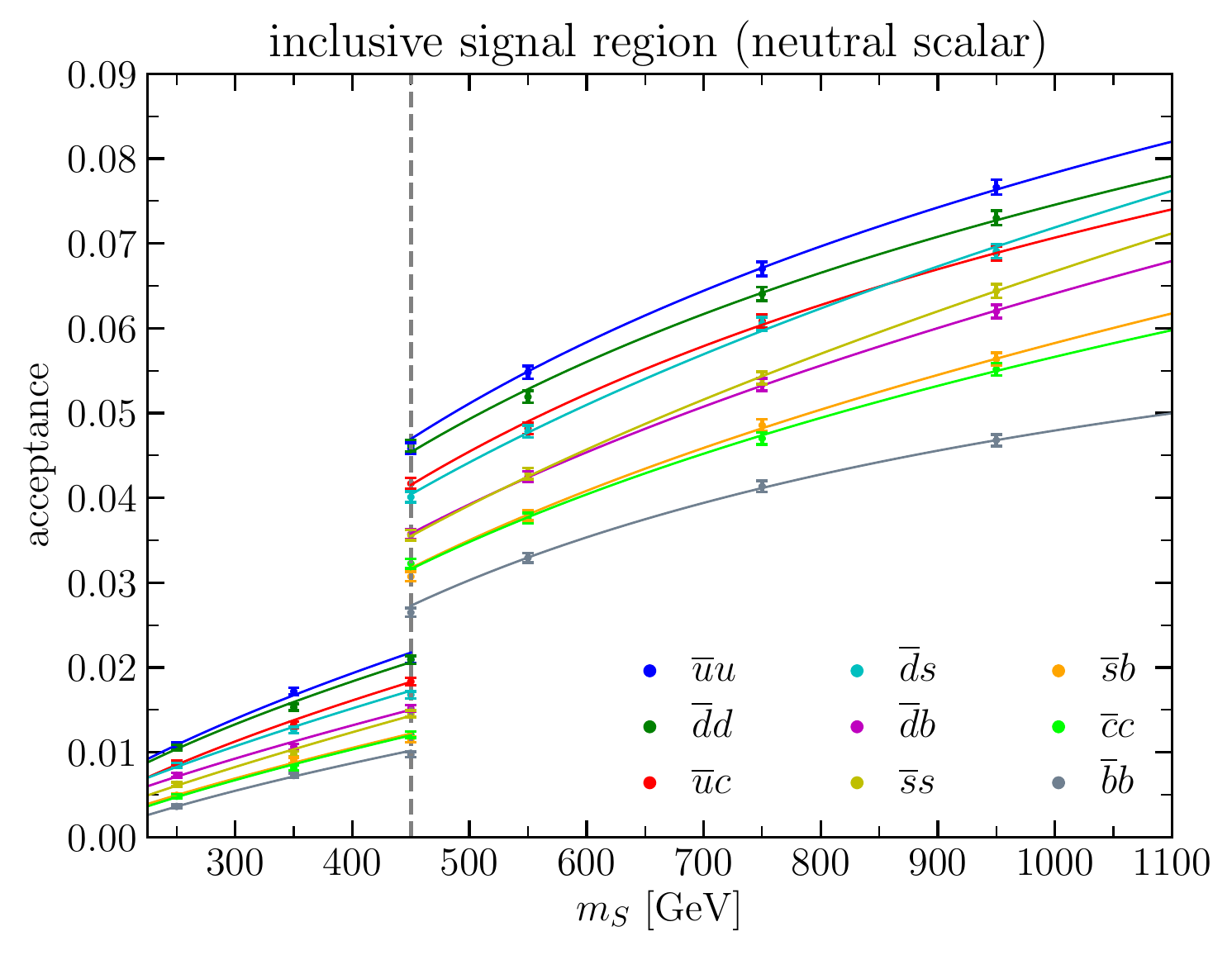}
  \end{minipage}
  \begin{minipage}{.48\textwidth}\centering
    \includegraphics[width=\textwidth]{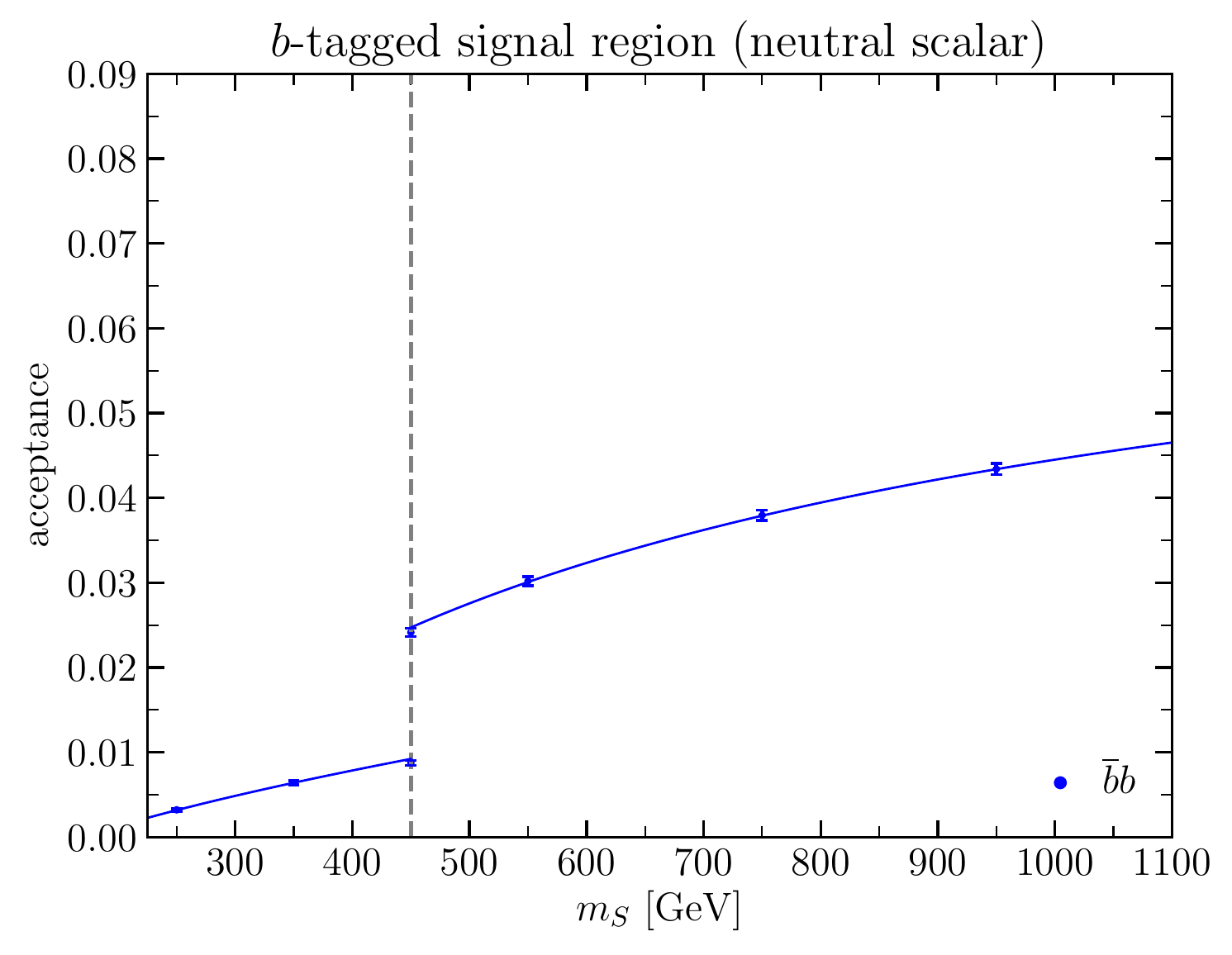}
  \end{minipage}
  \begin{minipage}{.48\textwidth}\centering
    \includegraphics[width=\textwidth]{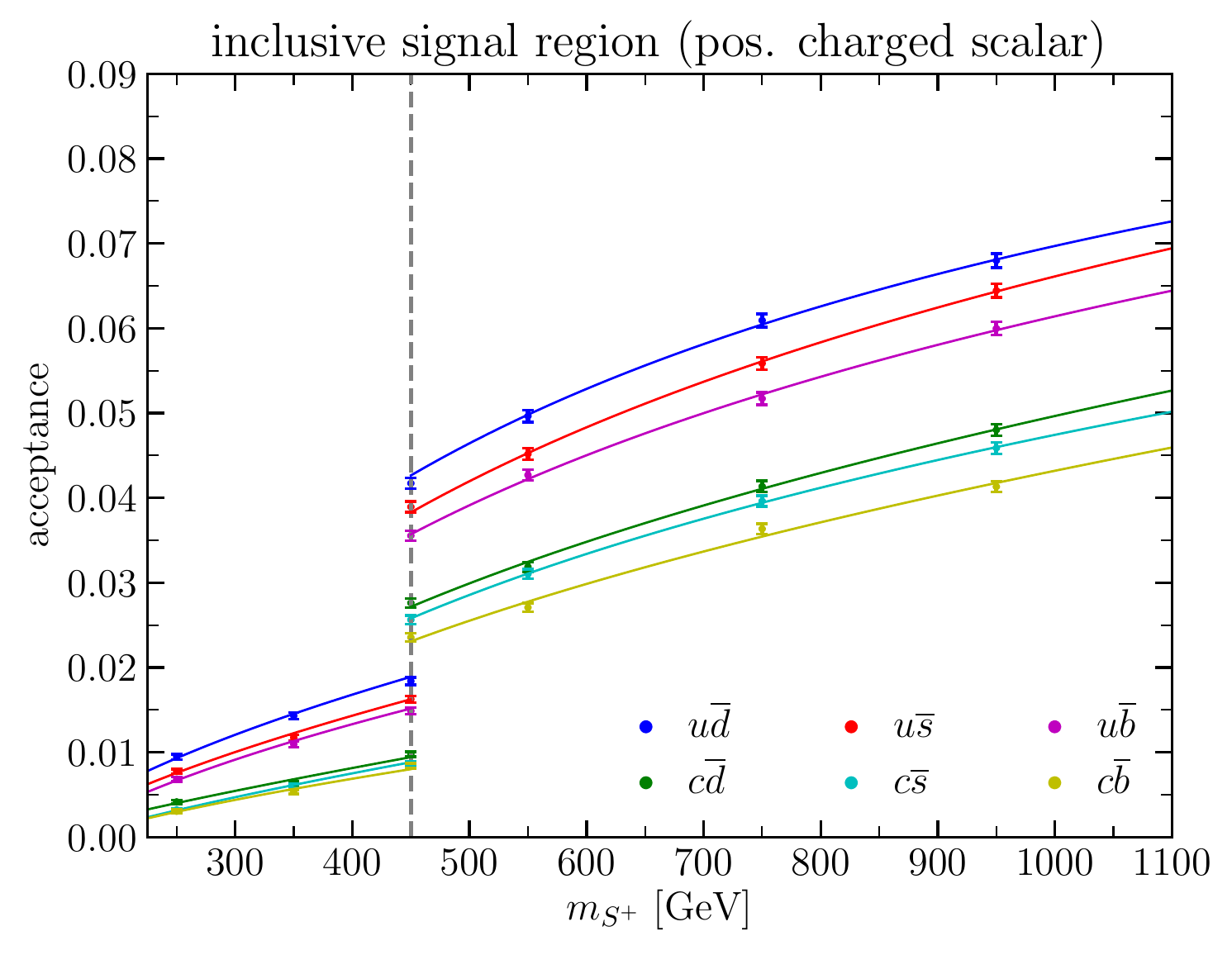}
  \end{minipage}
  \begin{minipage}{.48\textwidth}\centering
    \includegraphics[width=\textwidth]{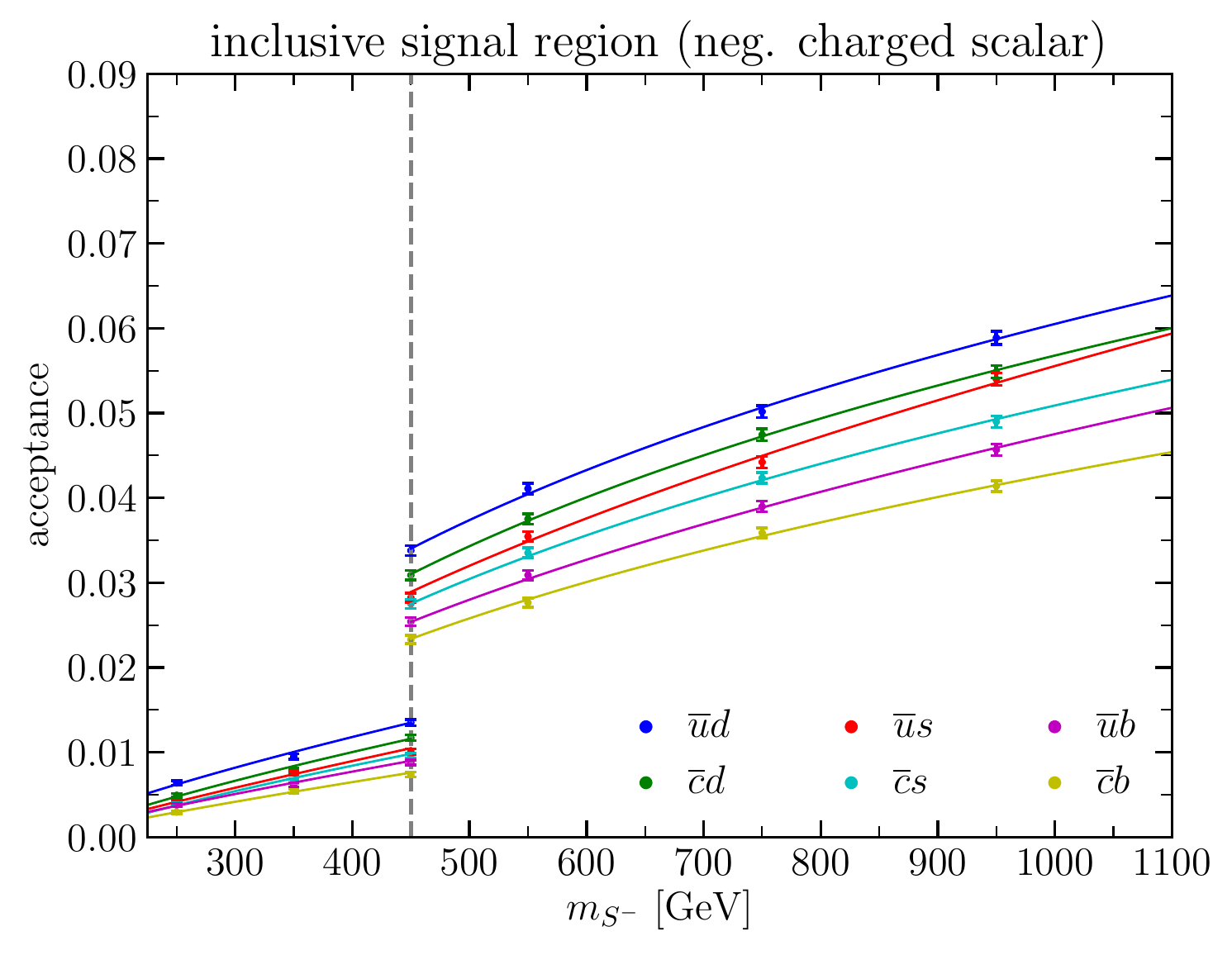}
  \end{minipage}
  \caption{The acceptance of the ATLAS search for a resonance with a di-jet final state produced in association with a photon. For this Figure, the scalar resonance is assumed to couple to only one type of quark pair (see legend). The points are used to denote parameter points for which explicit MC samples have been generated. The lines represent the corresponding fit functions. \textit{Upper left:} Acceptance of a neutral scalar resonance in the inclusive signal region. \textit{Upper right:} Acceptance of a neutral scalar resonance in the $b$-tagged signal region. \textit{Bottom:} Acceptances of a charged scalar resonance in the inclusive signal region.}%
  \label{fig:dijet_ISR_acceptances}
\end{figure}

We display some of the resulting acceptance fits in \cref{fig:dijet_ISR_acceptances}. For this Figure, we assume that the resonance couples to only one type of quark pairs (as shown in the plot legend) corresponding to only one coupling in \cref{eq:neutral_scalar_model} being non-zero. Note, however, that our implementation also fully supports the situation in which the initial-state quark pair is different from the final-state quark pair. The vertical dashed lines at $M_S = 450\gev$ mark the transition from the single-photon to the combined trigger. The markers denote points for which we produced MC samples. The uncertainty bands denote the corresponding statistical uncertainty. The colored lines represent the derived fit functions.

In the upper left panel of \cref{fig:dijet_ISR_acceptances}, the acceptances of a neutral scalar resonance in the inclusive signal region are shown in dependence of the resonance mass. While the acceptance is relatively low ($\sim 1-2\%$) in the single-photon-trigger region, it increase to up to $\sim 8\%$ in the combined-trigger region. As a consequence of the different PDFs for different quark flavors, the acceptance for $\bar u u$-initiated neutral scalar production (blue curve) is up to $\sim 3\%$ higher than e.g.\ the acceptance of the $\bar b b$-initiated channel (gray curve).

In the upper right panel of \cref{fig:dijet_ISR_acceptances}, the acceptance of a neutral scalar resonance in the $b$-tagged signal region is shown. As a consequence of the $b$ tagging, the acceptance is non-zero only for a $b\bar{b}$ final state. Due to the tighter cuts on the pseudo-rapidity of the jets, the acceptance is slightly reduced in comparison to the one of the $\bar b b$-initiated channel in the upper left plot of \cref{fig:dijet_ISR_acceptances}.

The acceptances for a charged scalar resonance in the inclusive signal region are shown in the bottom panels of \cref{fig:dijet_ISR_acceptances}. While the overall behavior closely resembles the behavior of the acceptances for a neutral scalar resonance, the overall acceptance values are reduced by $\sim 1-2\%$. This a consequence of the cuts being optimized for a neutral resonance not taking into account the possibility of the photon being radiated by the resonance itself (see right panel of \cref{fig:dijet_ISR_feynman}). Since the charged resonance can not decay to two bottom quarks, it does not contribute to the $b$-tagged signal region. Due to charge asymmetry of the proton's quark content, the acceptances for a negatively charged scalar (see bottom right panel of \cref{fig:dijet_ISR_acceptances}) are slightly smaller than the acceptances of for a positively charged scalar (see bottom left panel of \cref{fig:dijet_ISR_acceptances}).

As already mentioned above, we require the mean invariant mass for all events with a $m_{jj}$ value between $0.8 M$ and $1.2 M$, where $M$ is the signal mass, as an additional input for applying the limits for a Gaussian resonance derived in \ccite{Aaboud:2019zxd}. Following the same procedure as for the acceptance, we fit this mean mass as a function of the initial-state quarks and the signal mass. The derived formulas are implemented into \HiBo and used to apply the Gaussian limits presented in \ccite{Aaboud:2019zxd}. For reference, we list them in \cref{sec:fits}.

For a particle with different channels contributing, we add the cross-section values for the individual channels multiplied by the respective acceptances.


\section{Model applications}%
\label{sec:results}

In this Section, we discuss four model applications for the \HiBo extensions discussed in \cref{sec:searches}: constraints on a scalar particle mediating between SM and DM particles, constraints on BSM Higgs bosons with a non-SM like Yukawa coupling structure within a 2HDM framework, and constraints on the sneutrino and slepton sectors in the $R$-parity violating MSSM\@. We want to emphasize that the presented applications are exemplary applications and that the implemented searches can also be used to constrain other models.


\subsection{Scalar Dark Matter portal model}

As a first exemplary application for the extensions of \HiBo discussed in \cref{sec:searches}, we consider a scalar DM portal model, in which a neutral scalar mediates interactions between the SM particles and the DM particles. More concretely, we focus on a model which was used in \ccite{Aaboud:2017phn} for the interpretation of mono-jet plus missing transverse energy searches. In this model, a pseudo-scalar mediator directly couples to all quarks (except the top quark) with an universal interaction strength $g_q$. In addition, it couples to fermionic dark matter with the interaction strength $g_\chi$. For simplicity, we assume that the mediator does not couple to any other particles.

We choose this model as an example in order to highlight the interplay of mono-jet searches --- as performed in \ccite{Aaboud:2017phn} --- and di-jet searches --- as discussed in \cref{sec:searches}. Since a recasting of the mono-jet search of \ccite{Aaboud:2017phn} would be beyond the scope of the current paper, we choose the same DM mass as in Fig.~7a of \ccite{Aaboud:2017phn} (i.e.\ 1~GeV).

\begin{figure}
    \centering
    \includegraphics[width=.6\textwidth]{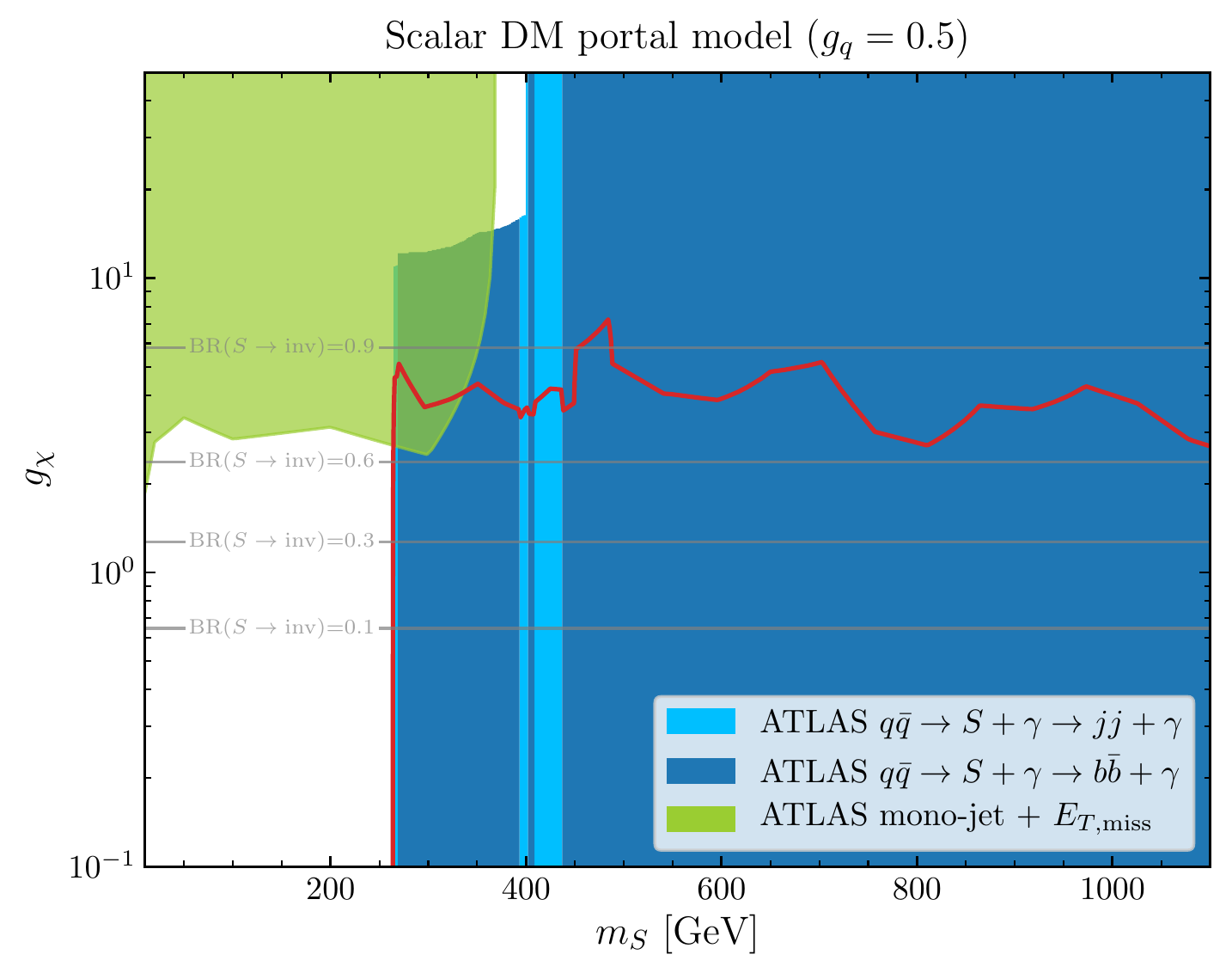}
    \caption{Collider constraints on scalar dark matter portal model in the $(m_S, g_\chi)$ parameter plane for $g_q = 0.5$. The blue coloured areas indicate which di-jet search is expected to be most sensitive. The green colored area is excluded by the ATLAS mono-jet plus missing energy search of \ccite{Aaboud:2017phn}. The area below the red curve is excluded by the ATLAS di-jet plus photon search~\cite{Aaboud:2019zxd}. The gray contours indicate the branching ratio of the scalar mediator $S$ into invisible particles.}%
    \label{fig:simpDM_model}
\end{figure}

This choice allows us to display the mono-jet limits of \ccite{Aaboud:2017phn} and the di-jet limits as implemented in \HiBo in the same plot (see \cref{fig:simpDM_model}). In this plot, the collider constraints are shown in the $(m_S, g_\chi)$ parameter plane, where $m_S$ is the mediator mass. The coupling of the mediator to quarks, $g_q$, is set to $0.5$. While the mediator is always produced via a di-quark initial state controlled by the coupling $g_q$, the relative strength of its decays into two quarks and two DM particles depends on the relative size of $g_\chi$ and $g_q$. The resulting branching ratio into DM particles is shown as gray contours in \cref{fig:simpDM_model}. While this branching ratio is below 10\% for $g_\chi\lesssim 0.65$, it lies above 90\% for $g_\chi\gtrsim 6$.

Correspondingly, the mono-jet plus missing transverse momentum search of \ccite{Aaboud:2017phn} limit (light green area) excludes the area of $g_\chi \gtrsim 2$ and $m_S \lesssim 370\gev$. The sensitivity of this searches decreases rapidly for higher mediator masses. As a complementary constraint, the di-jet plus photon search of \ccite{Aaboud:2019zxd} excludes the area below the red curve --- i.e., the region of $g_\chi \lesssim 3$ for mediator masses between $225\gev$ and $1100\gev$. The light (dark) blue regions indicate the area in which the inclusive ($b$-tagging) signal region of \ccite{Aaboud:2019zxd} is most sensitive.\footnote{For statistical consistency, \HiBo only applies the limit that yields with the strongest expected exclusion limit for each BSM particle and parameter point. See \ccite{Bechtle:2020pkv} for details.}

The results shown in \cref{fig:simpDM_model} demonstrate the complementary of mono- and di-jet searches for new resonances and the importance to provide a unified interpretation framework allowing to explore this complementarity.


\subsection{Two-Higgs-Doublet model}

One model where the impact of di-$b$-jet searches is known to be large is the Two-Higgs-Doublet Model (2HDM) with a flipped Yukawa sector. For a flipped Yukawa sector, the effective fermionic couplings of the non-SM-like neutral Higgs bosons $H$ and $A$ normalized to the respective couplings of the SM-like Higgs boson scale as
\begin{align}
    c(H/A u\bar{u})     & \propto \frac{1}{\tan\beta}\,, &
    c(H/A d\bar{d})     & \propto \tan\beta{}\,,           &
    c(H/A \ell^+\ell^-) & \propto \frac{1}{\tan\beta}\,,
\end{align}
if the Higgs at 125~GeV becomes SM-like. These effective couplings are independent of the fermion generation.

To demonstrate the impact of di-jet searches, we choose a simplified 2HDM parameter plane, where we only consider the heavy \cp-even neutral scalar $H$.
We assume that $\sin(\beta-\alpha)=1$, which puts the light \cp-even neutral scalar $h$ into the exact alignment limit so that it can easily accommodate the SM-like properties measured for the Higgs at 125~GeV. We then vary $\tan\beta$ and $m_H$ through a large range and use \texttt{2HDMC}~\cite{Eriksson:2009ws} linked to \HiBo to obtain and test the model predictions for the different parameter values.\footnote{For this demonstration we set the masses of all other Higgs bosons to \SI{2}{\TeV} (this i.e.~implies that the custodial symmetry is restored and electroweak precision constraints are evaded) and ignore all constraints associated with those particles. This has no impact on the properties of $H$ we are interested in.}

\begin{figure}
    \centering
    \includegraphics[width=0.8\textwidth]{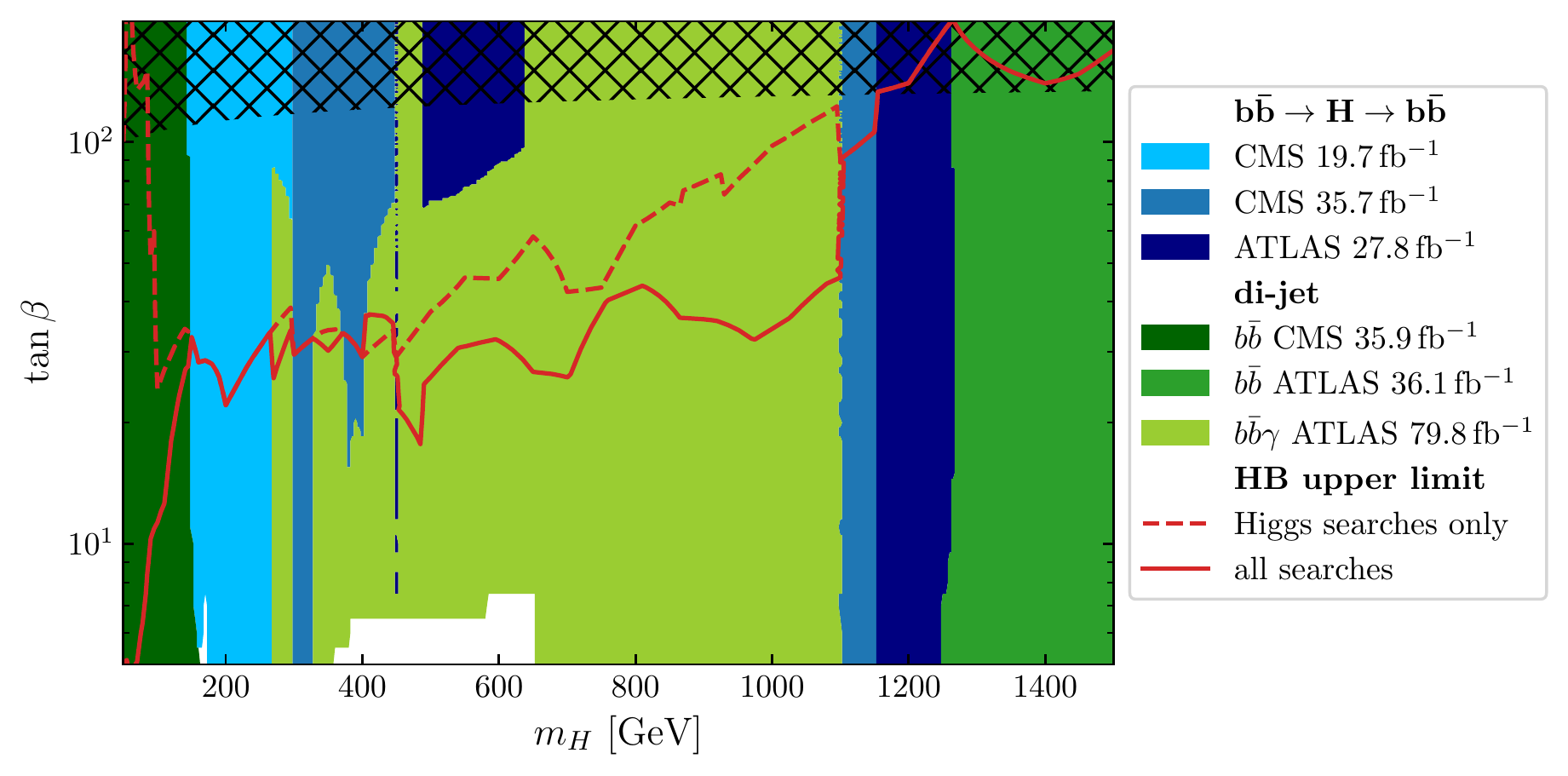}
    \caption{Limits from searches for the heavy Higgs boson $H$ in the flipped 2HDM in the ($m_H$,$\tan\beta$) plane. The color code indicates the most sensitive limit for each parameter point. The blue limits are Higgs searches in the $b\bar{b} \to H \to b\bar{b}$ channel~\cite{Khachatryan:2015tra,Sirunyan:2018taj, Aad:2019zwb}, while the green limits are newly implemented di-jet searches in the $b\bar{b}$~\cite{Aaboud:2018tqo,Sirunyan:2018ikr} and $b\bar{b}+\gamma$~\cite{Aaboud:2019zxd} channels. The resulting \texttt{HiggsBounds} upper limit on $\tan\beta$ is shown in red both for all searches (solid line) and for dedicated Higgs searches only (dashed line). The hatched region indicates where the narrow width approximation looses validity as $\Gamma_H^\text{tot}/m_H>0.25$.}%
    \label{fig:2HDM_bb}
\end{figure}

The results are shown in \cref{fig:2HDM_bb}. For each parameter point, the color code indicates which limit is most sensitive to the heavy BSM scalar $H$. The analyses indicated in blue are previously implemented Higgs searches targeting the $b\bar{b}\to H\to b\bar{b}$ channel~\cite{Khachatryan:2015tra,Sirunyan:2018taj, Aad:2019zwb}, while the green analyses are the newly implemented di-$b$-jet~\cite{Sirunyan:2018ikr,Aaboud:2018tqo} and di-$b$-jet+$\gamma$~\cite{Aaboud:2019zxd} searches. The red lines indicate the overall \HiBo \SI{95}{\%}~C.L. upper limit on $\tan\beta$.  The solid line includes all Higgs and di-jet searches, while the dashed line only includes dedicated Higgs searches. Finally, the hatched region indicates where $\Gamma_H^\text{tot} > m_H/4$, such that the narrow width approximation is no longer applicable.

Let us now consider the impact of the di-jet searches in \cref{fig:2HDM_bb} in detail. At $m_H<\SI{200}{\GeV}$ the low-mass di-$b$-jet search by CMS~\cite{Sirunyan:2018ikr} leads to a huge difference in the excluded parameter region. The only dedicated Higgs search in this region was \ccite{Khachatryan:2015tra}, which only covers $m_H\geq \SI{100}{\GeV}$. Below that mass, the only existing limits at large $\tan\beta$ came from LEP~\cite{Abdallah:2004wy} and the Tevatron~\cite{Abazov:2010ci}, leading to the weak exclusion of the dashed line. The newly implemented CMS di-$b$-jet limit, on the other hand, puts a very stringent upper bound on $\tan\beta$, especially for very low $m_H\lesssim\SI{70}{\GeV}$, where it excludes the region down to $\tan\beta\lesssim 6$.

For intermediate mass values, the exclusion is dominated by the newly implemented di-$b$-jet+$\gamma$ limit~\cite{Aaboud:2019zxd}. For $m_H\lesssim\SI{500}{\GeV}$ the limit is comparable to the dedicated CMS Higgs search~\cite{Sirunyan:2018taj}. Which of the two is the most sensitive changes based on the precise mass and --- due to the width dependence of the $b\bar{b}+\gamma$ limit --- $\tan\beta$ value. For larger masses, the $b\bar{b}+\gamma$ limit becomes more dominant, with a significantly stronger resulting limit on $\tan\beta$. On the one-hand, this behavior is not unexpected, since the $b\bar{b}+\gamma$ analysis uses $\sim\SI{80}{\fb^{-1}}$ of data, while all other analyses use at most $\sim\SI{36}{\fb^{-1}}$. However, especially for larger masses, the $b\bar{b}+\gamma$ search also benefits from the $H+\gamma$ production cross section, which falls off less rapidly with increasing Higgs mass.

Out of the dedicated Higgs searches, \ccite{Aad:2019zwb} covers the highest masses, with a limit set up until $m_H=\SI{1.4}{\TeV}$. The limit arising from the di-jet search \ccite{Aaboud:2018tqo} is substantially stronger already from $m_H\gtrsim\SI{1.3}{\TeV}$. However, both limits only exclude parameter points in the hatched region, where the narrow width approximation is no longer applicable. Therefore, at very high masses even with the newly implemented searches no physically meaningful limit can be set.

\subsection{\texorpdfstring{$R$}{R}-parity violating MSSM}

In the MSSM, the couplings of the scalar fermions are fixed by a small set of parameters. As a consequence of $R$ parity, these bosons always decay into a final state containing missing transverse energy. If $R$ parity is violated, however, the $R$ parity violating couplings of scalar fermions are essentially free parameters. Within the superpotential of RPV SUSY one can find the terms,
\begin{eqnarray}
\mathcal{W}_{RPV} \supset \epsilon_{ab}\left[\frac{1}{2}\lambda_{ijk}L_i^a L^b_j\bar{E}_k + \lambda'_{ijk}L_i^a Q_j^b\bar{D}_k\right].
\end{eqnarray}
Thus, the couplings of the sneutrinos and the sleptons are controlled by the parameters $\lambda_{ijk}$ --- coupling a left-handed slepton to two leptons ---, and $\lambda^\prime_{ijk}$ --- coupling a left-handed slepton to two quarks, where $i$, $j$, and $k$ are generation indices.

For a non-zero $\lambda^\prime$ coupling, the respective sneutrino and slepton can be produced at the LHC via a di-quark initial state. Moreover, the presence of this coupling allows the decay into a di-jet final state. For a non-zero $\lambda$ coupling, the sneutrino can decay into two charged leptons, whereas the slepton can decay to a neutrino and a charged lepton.

\begin{figure}
	\centering
	\includegraphics[width=.45\textwidth]{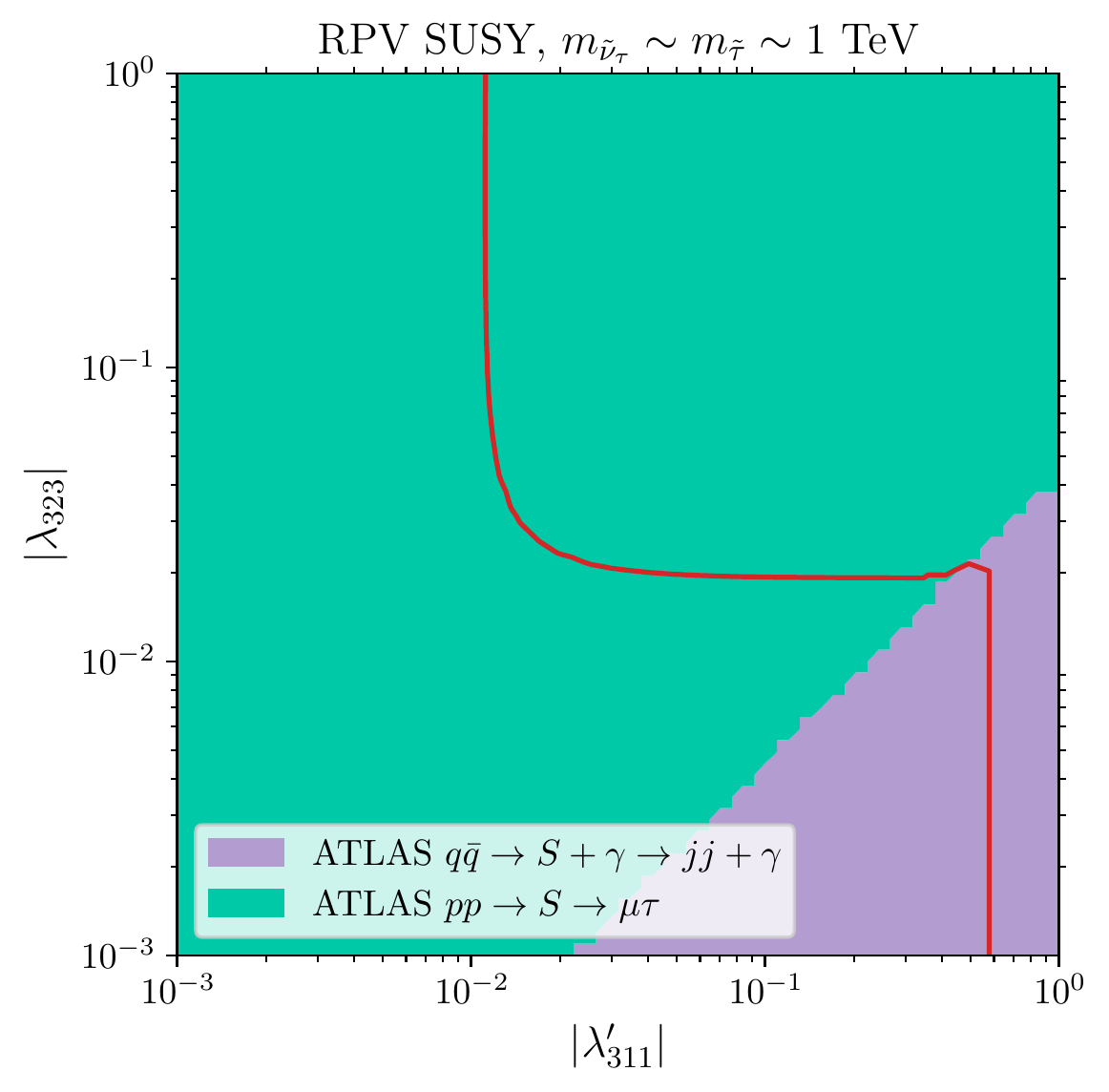}
	\includegraphics[width=.45\textwidth]{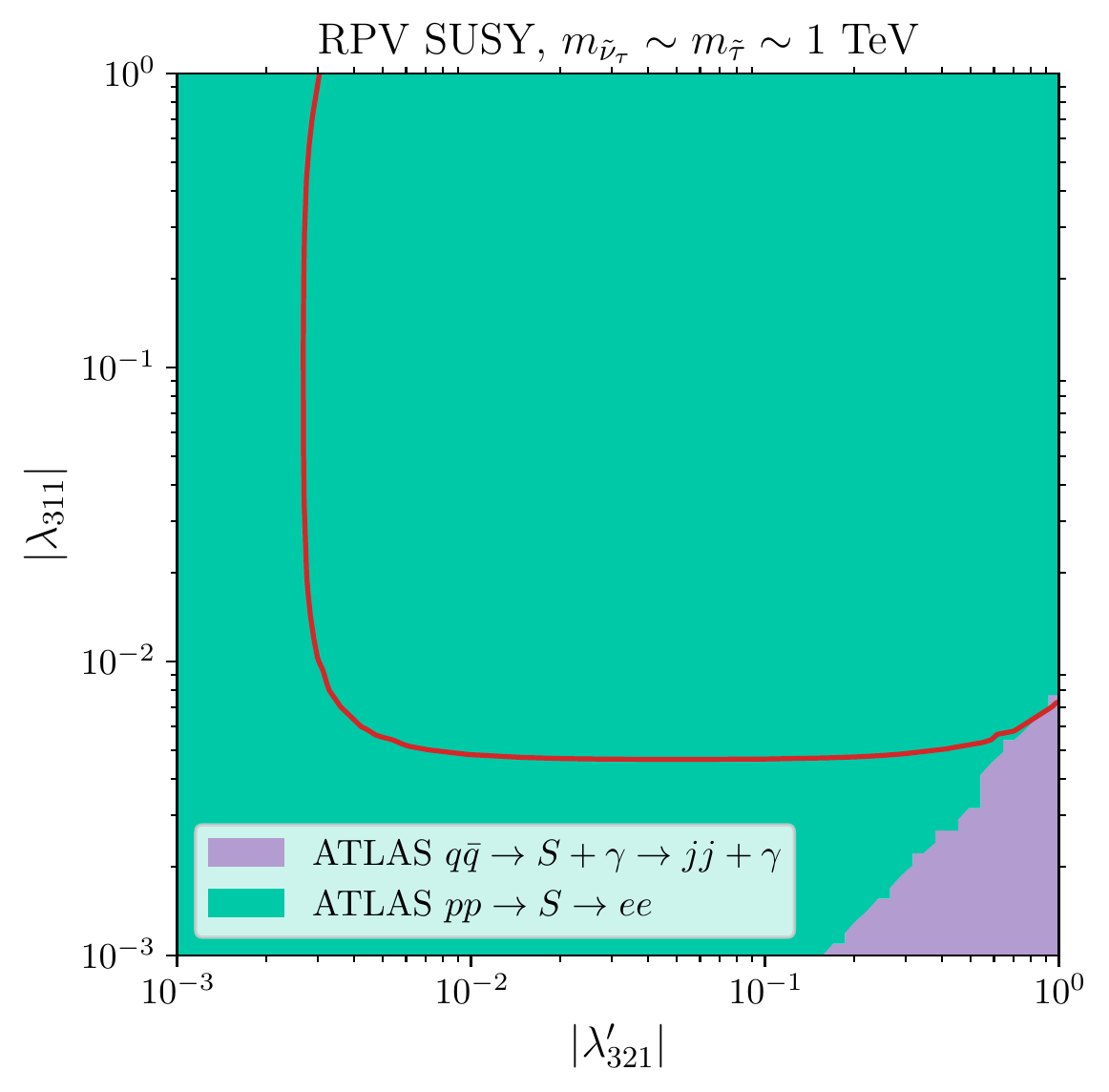}\\
	\includegraphics[width=.45\textwidth]{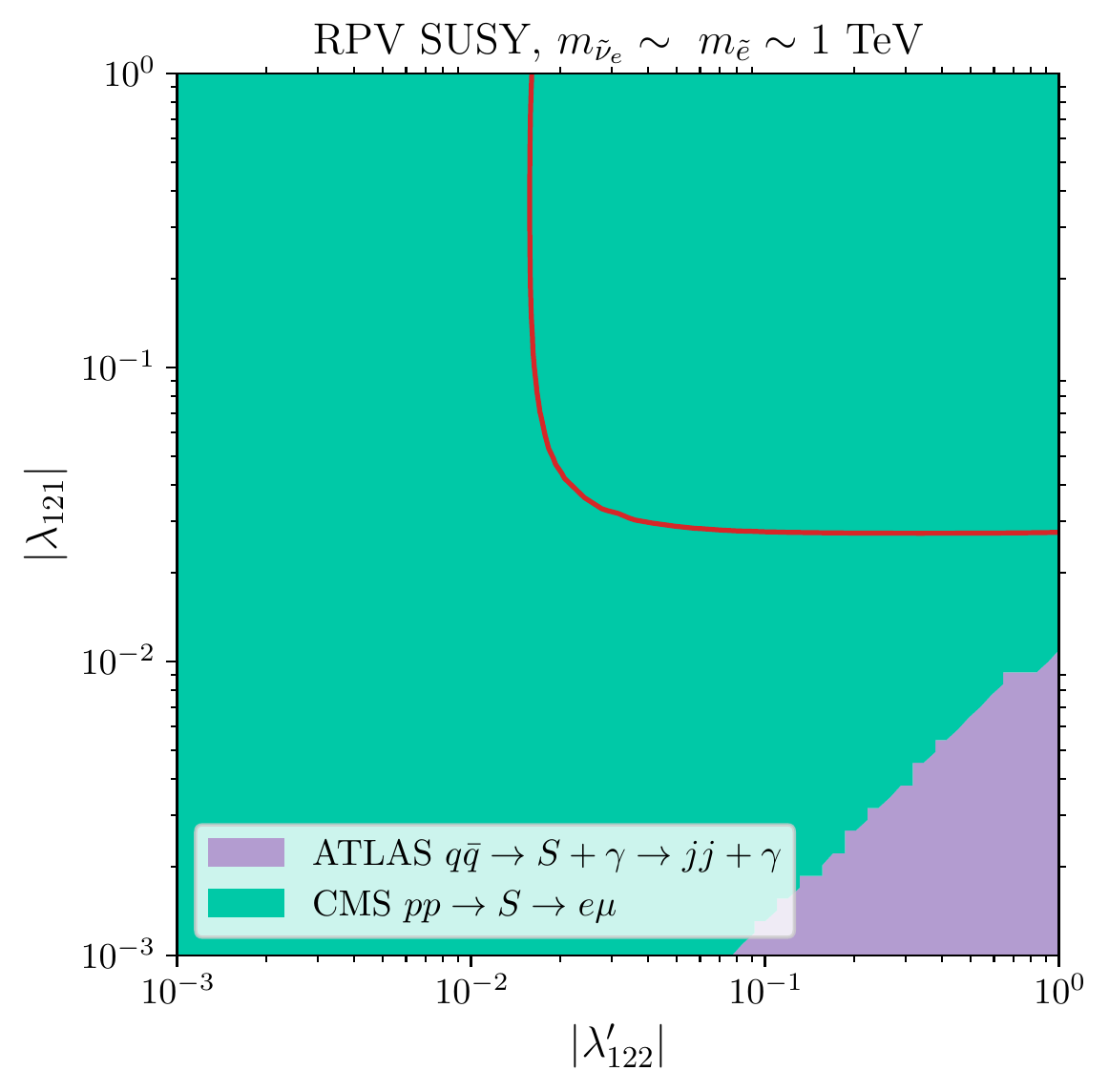}
	\includegraphics[width=.45\textwidth]{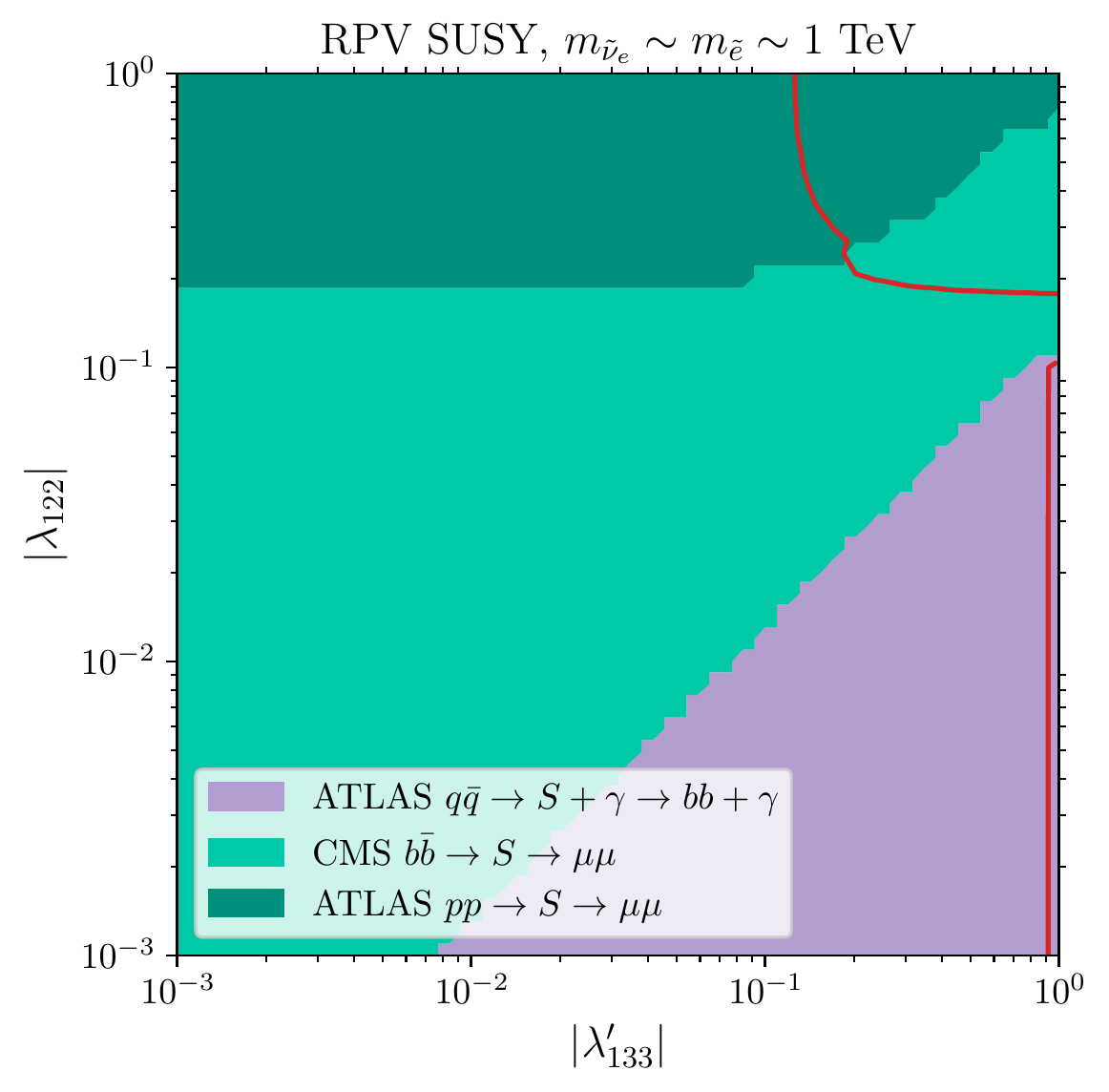}
	\caption{Limits from searches for a scalar resonance decaying to two jets or two leptons in the $R$-parity violating MSSM\@. The color code indicates the most sensitive limit for each parameter point. The red line indicates the exclusion limit set by \HiBo.}%
	\label{fig:RPV}
\end{figure}

In \cref{fig:RPV}, we display four exemplary parameter scans in which the two couplings $\lambda_{ijk} = -\lambda_{jik}$ and $\lambda_{ijk}^\prime$ are varied. We choose four scenarios that illustrate different production channels and phenomenological signatures that could be covered by existing searches. The top left plot in \cref{fig:RPV} we performed a scan over the two couplings $\lambda_{323} = -\lambda_{233}$ and $\lambda_{311}^\prime$. This means that the $\tau$ sneutrino, whose mass we choose to be at $\sim 1\tev$,\footnote{We fix the corresponding soft SUSY-breaking parameter to be equal $1\tev$. The additional contributions to the tau sneutrino and left-handed stau masses cause small deviations from $1\tev$ of $\sim1\gev$. All other SUSY particles are decoupled.} can be produced via a $d\bar d$ initial state and decays either to two jets or a muon and a tau lepton. The associated left-handed stau can be produced via a $ud$ initial state and decays into two jets or the leptonic channel, $\mu\nu_\tau$.\footnote{In all of the RPV SUSY scenarios, the stau and the sneutrino share the di-jet decay mode and are very close in mass --- within the estimated mass resolution of any di-jet search --- through the entire parameter plane. Therefore, \HiBo treats them as unresolved in the di-jet channel and sums their predicted rates when comparing to the di-jet limits, see \ccite{Bechtle:2013wla} for more details.}

Two of the newly implemented searches discussed in \cref{sec:searches} are sensitive to this signature: the ATLAS $pp\to S \to \mu\tau$ resonance search~\cite{Aaboud:2018jff} and the ATLAS di-jet plus photon search~\cite{Aaboud:2019zxd}. While the $\mu\tau$ search excludes the region of $|\lambda_{311}^\prime| \gtrsim 10^{-2}$ and $|\lambda_{323}| \gtrsim 2\cdot 10^{-2}$, the di-jet plus photon search allows to set an upper limit on $|\lambda_{311}^\prime|$ of $\sim 0.6$ showing nicely the complementarity between the different search channels. Note that here and in all the other scenarios, both the sneutrino and the --- almost mass degenerate --- associated slepton contribute to the di-jet rates. This is what makes the limits from di-jet final states competitive to the leptonic ones.

The top right plot of \cref{fig:RPV} shows the result of the scan of the two couplings $\lambda_{311} = -\lambda_{131}$ and $\lambda_{321}^\prime$. In this case the tau sneutrino can decay into a pair of jets or a pair of electrons while its production is caused by a $s\bar{d}\, (\bar{s}d)$ initial state. The stau for this configuration is produced mainly by $\bar{u}d\,(u\bar{d})$ initial state and its decay consists of a pair of jets or the pair $e^+e^-$. In this case the two most significant searches are the ATLAS di-lepton resonance search~\cite{Aad:2019fac} and the ATLAS di-jet plus photon search~\cite{Aaboud:2019zxd}. The highly sensitive $e^+e^-$ search leads to stringent constraints down to $|\lambda_{321}^\prime| \gtrsim 3\cdot 10^{-3}$ and $|\lambda_{311}| \gtrsim 5\cdot 10^{-3}$. As a consequence, the complementary di-jet plus photon search is only significant for large values of $|\lambda_{321}^\prime|$ and small values of $|\lambda_{311}|$.

The third scenario is shown in the bottom left plot of \cref{fig:RPV}, in which we scan over $\lambda_{121} = -\lambda_{211}$ and $\lambda_{122}^\prime$. In this case the lightest scalar fermions are the electron sneutrino and the selectron. The electron sneutrino can be produced in this case by a $\bar{s}s$ initial state and decays into a pair of jets or a muon and an electron. In this case the selectron, that is produced by the $c\bar{s}\,(\bar{c}s)$ initial state, decays into a pair of jets or the pair $e\mu$. The two most sensitive searches in this scenario are the ATLAS di-jet plus photon search~\cite{Aaboud:2019zxd} and the CMS $pp\to S \to e\mu$ resonance search~\cite{Sirunyan:2018zhy}. In this case the electron muon search can set the next bounds on the parameters $|\lambda_{122}^\prime| \gtrsim 1.5\cdot 10^{-2}$ and $|\lambda_{121}| \gtrsim 3\cdot 10^{-3}$.

The last scenario (bottom right plot of \cref{fig:RPV}) corresponds to a scan of the parameters $\lambda_{122} = -\lambda_{211}$ and $\lambda_{133}^\prime$. In this case the electron sneutrino decays into a pair of b-jets or two muons while it could be produced via a $\bar{b}b$ pair. The selectron that decays into a pair of jets or the pair $\mu\nu_\mu$, can be produced by the initial state $cb$. In this scenario, the three most sensitive searches are:
the newly implemented ATLAS two $b$-jets plus photon search~\cite{Aaboud:2019zxd} and ATLAS $pp\to S\to \mu\mu$ search~\cite{Aad:2019fac}, and the previously implemented CMS $\bar{b}b\to S\to \mu\mu$ search~\cite{Sirunyan:2019tkw}. The CMS muon search is able to constrain $|\lambda_{122}| \gtrsim 1.5\cdot 10^{-1}$ while the ATLAS di-muon search constrains $|\lambda_{133}^\prime| \gtrsim 1.3\cdot 10^{-1}$. It is interesting to see here why both searches cover different areas in the $(\lambda, \lambda^\prime)$ parameter space. The CMS $\bar{b}b\to S\to \mu\mu$ search~\cite{Sirunyan:2019tkw} that is sensitive to small values of $|\lambda_{122}|$ sets bounds on resonances up to a mass of $1\tev$. Higher values of $|\lambda_{122}|$ make the physical mass of the sneutrino slightly greater than $1\tev$ which makes this search insensitive. When we reach this threshold the ATLAS $pp\to S\to \mu\mu$ search~\cite{Aad:2019fac}, which covers higher resonance masses, becomes the most sensitive search. In this scenario we can see how the newly implemented searches are complementary to the previous ones.

As we have shown in the four cases appearing in \cref{fig:RPV} the newly implemented searches are quite sensitive and cover substantial parts of the parameter space of couplings. Our extension of \HiBo thus enables a deep exploration of the sneutrino/slepton sector of the RPV MSSM\@.


\section{Conclusions}%
\label{sec:conclusions}

BSM scalars are a well-motivated target for LHC searches. Normally, new scalar states are anticipated to have coupling structures similar to the SM Higgs boson. Many examples for BSM theories containing new scalar states exists, however, for which this is not the case. In the present article, we presented an extension of the code \HiBo allowing to test such scalars with a general coupling structure.

Besides extending the \HiBo input routines, we also implemented ATLAS and CMS di-lepton and di-jet searches into \HiBo based upon generic scalar models. While the implementation of the di-lepton searches is straightforward, we have developed a new approach to implement limits with a non-trivial dependence on the involved couplings (as it is the case for the implemented di-jet limits): By recasting the analysis results and tabulating the acceptances and efficiencies as a function of all involved couplings in a generic scalar model, we are able to evaluate limits for arbitrary BSM scalar models without the need to run Monte-Carlo simulations for every parameter point. While this approach can be applied to existing searches by the means of recasting, we encourage the experimental collaborations to directly provide the relevant acceptance and efficiency functions in a sufficiently general simplified model, e.g.\ in our generic scalar models. This would significantly improve the applicability of the experimental results without the need to re-implement the analysis chain. Moreover, this would also allow for more precise limit setting, since it is normally only possible to recreate the experimental analysis chain in an approximate manner.

Evaluating the acceptances and efficiencies in our generic scalar models also allows us to judge the impact of some commonly used approximations when no applicable model interpretations are available and recasting every parameter point is not feasible. As expected, we have found that the acceptances for a scalar can differ significantly from those of a spin-1 boson, such that applying $Z'$ or $W'$ limits to scalar particles is an extremely crude approximation. Additionally, we observed that the acceptances for a scalar can vary significantly depending on the specific initial and final state, such that scalar interpretations assuming flavor-universal couplings only give a rough estimate of the true acceptances if flavor universality is not fulfilled. While we only investigated scalar particles in this study, this conclusion should be equally valid for spin-1 particles --- e.g.\ in non-flavor-universal $Z'$ models --- and we advise caution when generalizing flavor universal limits to such scenarios.

We used the extended \HiBo version for several example applications. First, we demonstrated the complementarity of di-jet and mono-jet limits for constraining simplified DM models. As a second example, we investigated the impact of generic di-$b$-jet resonance searches on the flipped 2HDM excluding parameter regions which are not yet constrained by dedicated heavy Higgs searches. As a third example, we discussed constraints on sneutrino and slepton couplings in the $R$-parity violating MSSM\@. While the newly implemented di-lepton resonance searches set an upper limit on the product of sneutrino-lepton-lepton and sneutrino-quark-quark couplings, the di-jet limits allow to set an upper bound on the slepton- and sneutrino-quark-quark couplings regardless of the sneutrino-lepton-lepton couplings.

The presented extension of \HiBo provides a useful tool for constraining BSM model containing scalars with an ``exotic'' coupling structure. All of the features described in this paper are available from \HiBov{5.10.0} onwards, which is available at
\begin{center}
    \href{https:/gitlab.com/higgsbounds/higgsbounds}{\url{https://gitlab.com/higgsbounds/higgsbounds}}\,.
\end{center}
In the future, we plan to apply the presented approach for implementing more complex analyses into \HiBo also to other search targets like mono-jet plus missing energy final states.


\section*{Acknowledgments}
\sloppy{
We thank Caterina Doglioni, Antonio Boveia, Dan Guest, Gustavo Otero, Katherine Pachal for helpful discussions regarding the ATLAS di-jet plus photon search. We are also very grateful to Ayan Paul for helpful discussions. H.B., V.L. and T.S. are funded by the Deutsche Forschungsgemeinschaft (DFG, German Research Foundation) under Germany‘s Excellence Strategy -- EXC 2121 ``Quantum Universe'' – 390833306. The work of J.W. is funded by the Swedish Research Council, contract number 2016-0599.
}


\appendix

\section{ATLAS search for di-jet final state in association with initial-state photon radiation --- validation}%
\label{sec:dijetISR_valid}

In \cref{sec:dijetISR}, we describe the implementation of the ATLAS search for a di-jet resonance produced in association with an initial-state photon into the \MA framework. Here, we validate this implementation.

For the validation, we use the same model as employed in \ccite{Aaboud:2019zxd}. I.e., we generate MC events for a leptophobic $Z'$ resonance using the \textit{DMsimp\_s\_spin1} \texttt{UFO} model~\cite{Abercrombie:2015wmb}. As in \ccite{Aaboud:2019zxd}, the $Z'$ resonance is assumed to have only axial-vector couplings to quarks with a universal coupling strength $g_q$ (the coupling to top-quarks is, however, set to zero). All couplings to other particles are set to zero.

Using the same setup as described in \cref{sec:dijetISR}, we generate MC samples consisting out of $10^5$ events for two benchmark points (BPs):
\begin{enumerate}
  \item BP1: $m_{Z'} = 250\gev$ and $g_q = 0.2$,
  \item BP2: $m_{Z'} = 550\gev$ and $g_q = 0.2$.
\end{enumerate}
For these two benchmark points, we can compare our analysis to the ATLAS cut flow tables which are given in the auxiliary material of \ccite{Aaboud:2019zxd}.

\begin{table}
  \begin{center}
  \renewcommand{\arraystretch}{1.}
    \begin{tabular}{l||cc|ccc}
      & \multicolumn{2}{c|}{ATLAS BP1} & \multicolumn{3}{c}{\MA BP1} \\ \hline\hline
      & Events & $\varepsilon$ & Events & $\varepsilon$ & $\delta$ [\%]\\ \hline
      Initial                                 & 7802.5 & --  & 7802.5 & -- & -- \\
      Trigger + Cleaning                      & 222.6 $ \pm $ 2.4 & 0.03 & 193.8 $ \pm $ 3.8 & 0.03 & 12.9 \\
      $n_\text{jets} > 2$                     & 207.1 $ \pm $ 2.3 & 0.93 & 175.4 $ \pm $ 3.7 & 0.91 & 2.7 \\
      $n_\gamma \ge 1$                        & 176.7 $ \pm $ 2.1 & 0.85 & 162.1 $ \pm $ 3.5 & 0.92 & 8.3 \\
      Lead $\gamma$ $p_T$                     & 143.8 $ \pm $ 1.9 & 0.81 & 136.2 $ \pm $ 3.2 & 0.84 & 3.2 \\
      Lead jet $p_T$                          & 143.8 $ \pm $ 1.9 & 1.00 & 136.2 $ \pm $ 3.2 & 1.00 & 0.0 \\
      Sublead jet $p_T$                       & 143.8 $ \pm $ 1.9 & 1.00 & 136.2 $ \pm $ 3.2 & 1.00 & 0.0 \\
      $|y*| < 0.75$                           & 108.4 $ \pm $ 1.7 & 0.75 & 102.8 $ \pm $ 2.8 & 0.75 & 0.1 \\
      Minimum $m_{jj}$                        & 81.5 $ \pm $ 1.4 & 0.75 & 77.2 $ \pm $ 2.4 & 0.75 & 0.1 \\ \hline
      $b$-tagging                             & 12.7 $ \pm $ 0.6 & 0.09 & 10.1 $ \pm $ 0.9 & 0.13 & 16.3 \\
    \end{tabular}
    \caption{Search for a di-jet resonance produced in association with a photon: cut flows for BP1 comparing the \MA event analysis as described in \cref{sec:dijetISR} and the numbers provided by ATLAS\@. For BP1, the single-photon trigger is employed.}%
    \label{tab:dijetISR_validation_BP1}
  \end{center}
\end{table}

The comparison of the cut flow tables can be found in \cref{tab:dijetISR_validation_BP1,tab:dijetISR_validation_BP2}. This table shows the number of events remaining after each cut and the associated statistical uncertainty (corresponding to the uncertainty of the MC integration). We reconstructed the initial number of events by dividing the final number of events by the total acceptance values given in the auxiliary material of \ccite{Aaboud:2019zxd}.\footnote{The final event number after the $b$-tagging step deviates from the number given in the auxiliary material of \ccite{Aaboud:2019zxd}. The number given there also includes the efficiency of the $b$-tagging signal region.} The columns marked by $\epsilon$ denote the cut efficiencies. The column denoted by $\delta$ shows the deviation between the cut efficiencies of \ccite{Aaboud:2019zxd} with respect to the \MA implementation.

For BP1 employing the single-photon trigger (see \cref{tab:dijetISR_validation_BP1}), the cut efficiencies of our \MA implementation are quite close to the numbers from \ccite{Aaboud:2019zxd}. Only the cut efficiency of the initial trigger and cleaning step as well as the $b$-tagging cut deviate by more than 10\%. The numbers are, however, still well below the agreement level of under 30\% recommended in \ccite{Dumont:2014tja}. Also note that the total acceptance values (before and after the $b$ tagging), obtained by dividing the final number of events by the initial number of events, are very close with absolute deviations smaller than $0.1\%$.

\begin{table}
  \begin{center}
  \renewcommand{\arraystretch}{1.}
    \begin{tabular}{l||cc|ccc}
      & \multicolumn{2}{c|}{ATLAS BP2} & \multicolumn{3}{c}{\MA BP2} \\ \hline\hline
      & Events & $\varepsilon$ & Events & $\varepsilon$ & $\delta$ [\%]\\ \hline
      Initial                                 & 801.2 & --  & 801.2 & -- & -- \\
      Trigger + Cleaning                      & 101.0 $ \pm $ 0.8 & 0.13 & 114.1 $ \pm $ 0.9 & 0.14 & 13.0 \\
      $n_\text{jets} > 2$                     & 99.8 $ \pm $ 0.8 & 0.99 & 109.9 $ \pm $ 0.9 & 0.96 & 2.4 \\
      $n_\gamma \ge 1$                        & 90.9 $ \pm $ 0.8 & 0.91 & 85.7 $ \pm $ 0.8 & 0.78 & 14.4 \\
      Lead $\gamma$ $p_T$                     & 88.6 $ \pm $ 0.7 & 0.98 & 85.7 $ \pm $ 0.8 & 1.00 & 2.6 \\
      Lead jet $p_T$                          & 88.4 $ \pm $ 0.7 & 1.00 & 85.4 $ \pm $ 0.8 & 1.00 & 0.1 \\
      Sublead jet $p_T$                       & 83.8 $ \pm $ 0.7 & 0.95 & 79.4 $ \pm $ 0.8 & 0.93 & 1.9 \\
      $|y*| < 0.75$                           & 58.5 $ \pm $ 0.6 & 0.70 & 54.9 $ \pm $ 0.6 & 0.69 & 0.9 \\
      Minimum $m_{jj}$                        & 49.6 $ \pm $ 0.6 & 0.85 & 46.6 $ \pm $ 0.6 & 0.85 & 0.1 \\ \hline
      $b$-tagging                             & 8.4 $ \pm $ 0.2 & 0.09 & 7.0 $ \pm $ 0.2 & 0.15 & 11.5 \\
    \end{tabular}
    \caption{Search for a di-jet resonance produced in association with a photon: cut flows for BP1 comparing the \MA event analysis as described in \cref{sec:dijetISR} and the numbers provided by ATLAS\@. For BP2, the combined trigger is employed.}%
    \label{tab:dijetISR_validation_BP2}
  \end{center}
\end{table}

The agreement level is very similar for BP2 employing the combined trigger (see \cref{tab:dijetISR_validation_BP2}). While the cut efficiency of the number of photons cut deviates by more than 10\%, all deviations in the cut efficiencies are still well below 30\%. Also the total acceptance values deviate by less than $0.4\%$.

\begin{figure}
\begin{minipage}{.48\textwidth}\centering
\includegraphics[width=\textwidth]{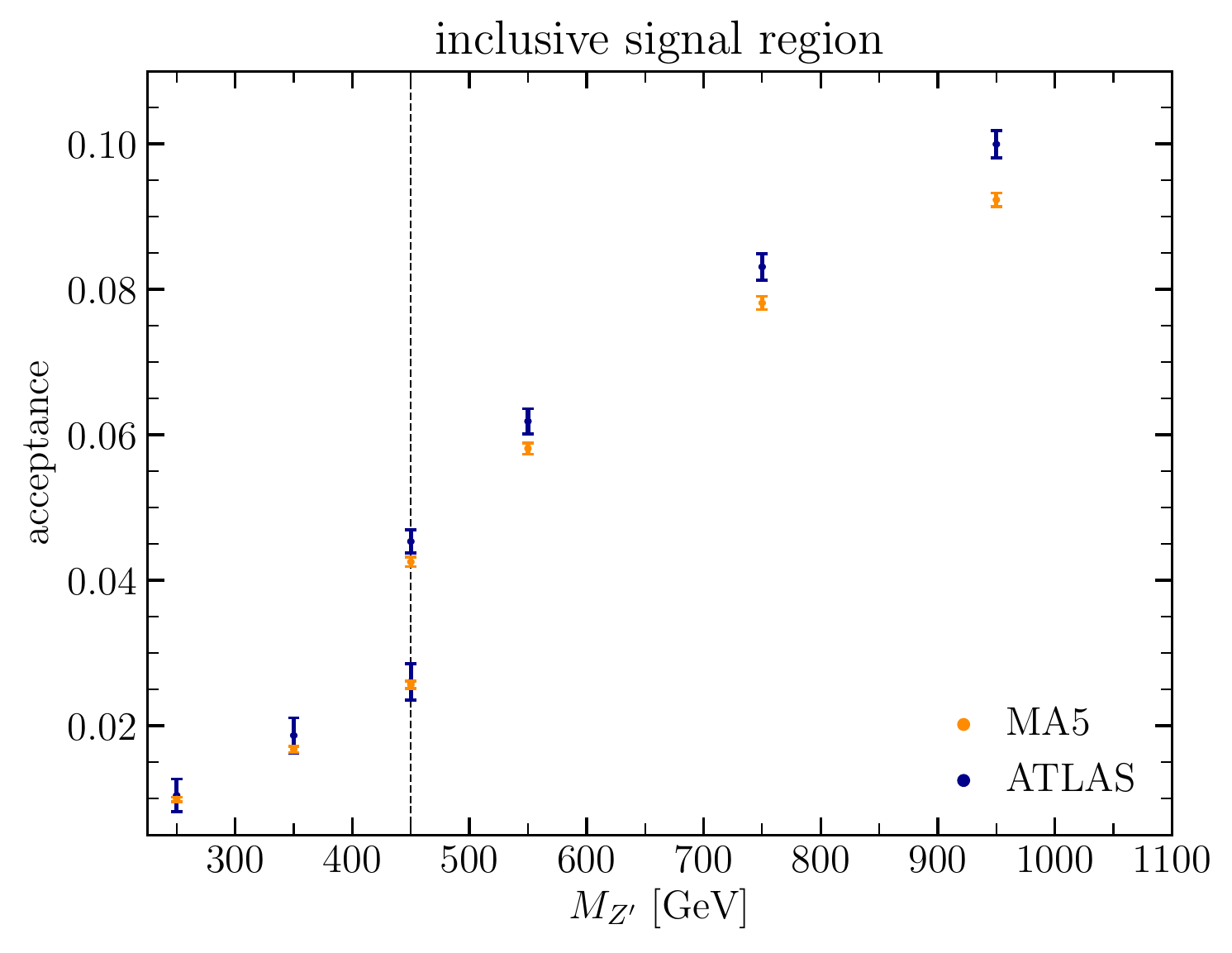}
\end{minipage}
\begin{minipage}{.48\textwidth}\centering
\includegraphics[width=\textwidth]{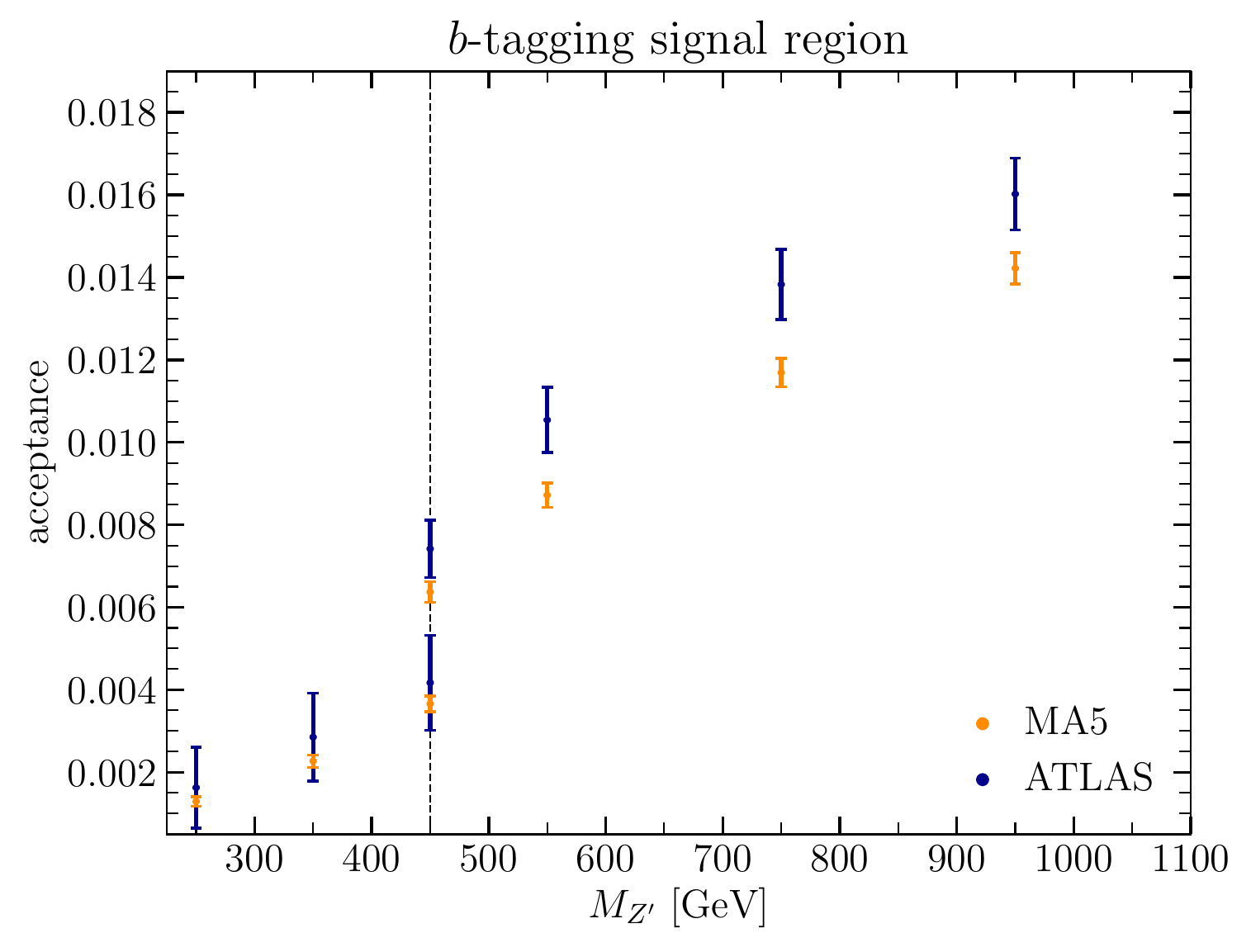}
\end{minipage}
\caption{Comparison of the acceptance values given in \ccite{Aaboud:2019zxd} to the values derived with our \MA implementation. \textit{Left:} inclusive signal region. \textit{Right:} $b$-tagged signal region.}%
\label{fig:Zprime_acceptances}
\end{figure}

As an additional validation, we compare the total acceptance values for different $M_{Z'}$ masses given in \ccite{Aaboud:2019zxd} to our implementation. This comparison is shown in \cref{fig:Zprime_acceptances}. In the left panel --- showing the acceptances for the inclusive signal region --- the agreement with the numbers given in \ccite{Aaboud:2019zxd} is very good for the single-photon trigger used for $M_{Z'} \le 450\gev$. For heavier $Z'$ masses, for which the combined trigger is used, the agreement is slightly worse with a maximum deviation of $\sim 1\%$. For the $b$-tagging signal region, shown in the right panel of \cref{fig:Zprime_acceptances}, a very similar behavior is visible. The overall size of the acceptances is, however, reduced by a factor $\sim 5$. The maximum deviation between the numbers given in \ccite{Aaboud:2019zxd} and our implementation amounts to $\sim 0.2\%$ for $M_{Z'} = 950\gev$.

Given this level of agreement, we consider our \MA implementation as validated.


\section{Fit formulas cross sections and efficiencies}%
\label{sec:fits}

In this section we list the fit formulas used in the implementation of the di-jet + photon search~\cite{Aaboud:2019zxd}.

\subsection{Cross section fits}
As described in \cref{sec:dijetISR_XS_fit}, we evaluated the  $pp \to \phi \gamma $ cross sections for each initial flavour separately and performed fits to this data. We use the functional form
\begin{equation}
    \frac{\sigma(g,m_S)}{\si{pb}} = g^2 \left(\frac{a}{m_S^3} + \frac{b}{m_S^2} + \frac{c}{m_S} + d + e\, m_S\right)\label{eq:cxnfit}
\end{equation}
for the cross section fits, where $g$ is the respective coupling as defined in \cref{eq:neutral_scalar_model,eq:charged_scalar_model} and $m_S$ is the particle mass in \si{GeV}. The values of the fit coefficients for the different initial state flavours and final state charges are given in \cref{tab:cxncoeffs}.

\begin{table}
    \sisetup{
        detect-mode,
        round-mode              = figures,
        round-precision         = 4,
        scientific-notation=true,
        table-format=-1.3e-1,
        group-digits=false,
    }
    \centering
    \begin{tabular}{l SSS[scientific-notation=false,table-format=-4.1,round-mode=places,round-precision=1]S[scientific-notation=false,table-format=-1.4,round-mode=places,round-precision=4]S}
        \toprule
                   & {$a$}        & {$b$}       & {$c$}      & {$d$}       & {$e$}            \\
                   & \si{\GeV^3}  & \si{\GeV^2} & \si{\GeV}  &             & \si{\GeV^{-1}}   \\
        \midrule
        $uu$       & -7.40884D07  & 4.49543D06  & 6390.61D0  & 3.69548D0   & -0.000795657     \\
        $dd$       & -2.60547D07  & 977832.D0   & -1895.61D0 & 1.6094D0    & -0.000534126D0   \\
        $cc$       & 3.63856D7    & 207310.D0   & -680.539D0 & 0.71179D0   & -0.000256867D0   \\
        $ss$       & 1.57635D07   & 61422.7D0   & -184.656D0 & 0.172986D0  & -0.0000559835D0  \\
        $bb$       & 567982.0D0   & 201805.0D0  & -448.806D0 & 0.406157D0  & -0.000136612D0   \\
        $uc$       & 5.65276D07   & 3.48162D06  & -6279.74D0 & 4.49248D0   & -0.00119019      \\
        $ds$       & 1.68853D07   & 699651.D0   & -1510.12D0 & 1.28013D0   & -0.00040275D0    \\
        $db$       & 8.13777D6    & 330773.0D0  & -711.803D0 & 0.607085D0  & -0.000192274D0   \\
        $sb$       & -3.90384D6   & 279829.0D0  & -646.173D0 & 0.613265D0  & -0.000215673D0   \\
        \midrule
        $u\bar{d}$ & - 1.73381D07 & 652027.0D0  & -1223.17D0 & 0.98188D0   & - 0.000307538D0  \\
        $c\bar{s}$ & + 5.3322D06  & 52331.9D0   & -136.529D0 & 0.125393D0  & - 0.0000409939   \\
        $u\bar{s}$ & + 3.71034D06 & 142254.0D0  & -351.939D0 & 0.335395D0  & - 0.000116551D0  \\
        $c\bar{d}$ & - 1.5052D07  & 457490.0D0  & -1006.88D0 & 0.917357D0  & - 0.000313157    \\
        $u\bar{b}$ & + 2.70981D06 & 72679.9D0   & -190.285D0 & 0.182345D0  & - 0.0000624154D0 \\
        $c\bar{b}$ & + 2.84796D06 & 28145.8D0   & -82.7795D0 & 0.0822505D0 & - 0.0000285173D0 \\
        \midrule
        $\bar{u}d$ & -471225D0    & 445930D0    & -247.076D0 & -0.216092D0 & 0.000173311D0    \\
        $\bar{c}s$ & 6.77824D6    & 34604.7D0   & -104.975D0 & 0.102024D0  & -0.0000347606D0  \\
        $\bar{u}s$ & 2.83736D6    & 296278D0    & -329.227D0 & 0.0805123D0 & 0.0000216609D0   \\
        $\bar{c}d$ & 2.87478D6    & 148855D0    & -395.107D0 & 0.39282D0   & - 0.000140121D0  \\
        $\bar{u}b$ & - 2.48801D7  & 396466D0    & -815.75D0  & 0.775128D0  & - 0.000288148D0  \\
        $\bar{c}b$ & + 3.8965D6   & + 19429.3D0 & -57.5982D0 & 0.0525333D0 & - 0.000016354D0  \\
        \bottomrule
    \end{tabular}
    \caption{Fit coefficients of \cref{eq:cxnfit} at the
        \SI[scientific-notation=false]{13}{\TeV} LHC\@. The top rows are for a
        neutral particle, the middle rows for a positively charged and the
        bottom rows for a negatively charged particle.}
    \label{tab:cxncoeffs}
\end{table}

\subsection{Acceptances and mass corrections}

As described in \cref{sec:dijetISR_acceptance_fit}, the acceptances and mass corrections are similarly fit to simulations for each initial or final state flavour and electric charge. The functional form used is
\begin{equation}
    \mathcal{A} = a + b\,\;\frac{m-\SI{225}{\GeV}}{\si{\GeV}} + c\; \log\frac{m}{\si{\GeV}}\label{eq:accfit}\,.
\end{equation}
For the preselection acceptance $\mathcal{A}_\text{initial}$ this is a function of the particle mass $m=m_S$ and depends on the initial state flavours with the coefficients for the two trigger categories shown in \cref{tab:acccoeffPre}. The final acceptance $\mathcal{A}_{m_{jj}}$ is a function of $m=m_S$ --- the mean mass after the $m_{jj}$ cut --- and depends on the final state flavours with the coefficients shown in \cref{tab:acccoeffPost}.

\begin{table}
    \sisetup{
        detect-mode,
        round-mode              = figures,
        round-precision         = 3,
        scientific-notation=true,
        table-format=-1.2e-1,
        group-digits=false,
    }
    \centering
    \begin{tabular}{l @{\hspace{1cm}}S[scientific-notation=false,table-format=-1.4,round-mode=places,round-precision=4]S[table-format=1.2e-1]S[table-format=1.2e-1]@{\hspace{1cm}}S[scientific-notation=false,table-format=-1.3]SS[scientific-notation=false,table-format=1.4]}
        \toprule
                   & \multicolumn{3}{c}{single photon trigger} & \multicolumn{3}{c}{combined trigger}                                                                                                       \\
                   & {$a$}                                     & {$b$}                                & {$c$}                   & {$a$}                  & {$b$}                   & {$c$}                  \\
        \midrule
        $uu$       & -0.07453818172065914D0                    & 3.824816674227349D-05                & 0.016692363267610276D0  & -0.3495397605469713D0  & -1.2642459929452903D-05 & 0.06945421537488733D0  \\
        $dd$       & -0.08365502599272057D0                    & 2.472432083149923D-05                & 0.018160770096650507D0  & -0.3471775210703532D0  & -2.5919492355969983D-05 & 0.06893383402208757D0  \\
        $cc$       & -0.03036038738244472D0                    & 4.221832611704063D-05                & 0.006949082950767061D0  & -0.25771965185274837D0 & 5.2868594897241616D-06  & 0.05060836725764018D0  \\
        $ss$       & -0.04243315863987981D0                    & 3.866314211122518D-05                & 0.009348295231225559D0  & -0.2161349881915616D0  & 2.0265589959806386D-05  & 0.04342174072764107D0  \\
        $bb$       & -0.02914949139483562D0                    & 3.419393921510185D-05                & 0.006807103486670389D0  & -0.2613043396316116D0  & -3.894324500526707D-06  & 0.051132326713863516D0 \\
        $uc$       & -0.0689894294152424D0                     & 3.8902627867545764D-05               & 0.015043024297551865D0  & -0.3485345843201749D0  & -1.223497402338667D-05  & 0.06839930939545714D0  \\
        $ds$       & -0.0453815009102044D0                     & 3.744108513321609D-05                & 0.010614770882040572D0  & -0.25385483138026416D0 & 6.894729090512683D-06   & 0.05128069341398131D0  \\
        $db$       & -0.048765522647684335D0                   & 3.456462028873385D-05                & 0.01114951745281092D0   & -0.2726412110713955D0  & 7.846779718611548D-07   & 0.054373087775165D0    \\
        $sb$       & -0.04967090498121169D0                    & 3.350800124436956D-05                & 0.010467748237181393D0  & -0.2784463150279004D0  & -2.3232981890858514D-06 & 0.054315795840985275D0 \\
        \midrule
        $u\bar{d}$ & -0.07335163824881155D0                    & 2.520750232361978D-05                & 0.01592191407492654D0   & -0.32901275609576014D0 & -2.329295062067327D-05  & 0.06498087881664942D0  \\
        $c\bar{s}$ & -0.029160671828213205D0                   & 2.8358926187903368D-05               & 0.006174353616870782D0  & -0.2138055657620705D0  & -5.777785130482657D-07  & 0.04162969122251766D0  \\
        $u\bar{s}$ & -0.06261671128216949D0                    & 2.7693711065568606D-05               & 0.01351445163517528D0   & -0.29351696001599786D0 & -1.115565824728008D-05  & 0.057883413664741076D0 \\
        $c\bar{d}$ & -0.014512733956164226D0                   & 3.2090057309420846D-05               & 0.0037939580782129917D0 & -0.19235684407120268D0 & 6.230537342199852D-06   & 0.038157270574518884D0 \\
        $u\bar{b}$ & -0.07217431189967927D0                    & 2.5143326400233142D-05               & 0.015194175446230258D0  & -0.3087878807658839D0  & -1.4223220997063607D-05 & 0.060621421727487405D0 \\
        $c\bar{b}$ & -0.032923747335909186D0                   & 2.375420069088394D-05                & 0.006937870226189841D0  & -0.17933973462875433D0 & 1.0897748362352643D-05  & 0.035417916916325634D0 \\
        \midrule
        $\bar{u}d$ & -0.04484663445244751D0                    & 2.917809069238506D-05                & 0.009845381304258653D0  & -0.25710823612127615D0 & -5.034398780010226D-06  & 0.05058736718447029D0  \\
        $\bar{c}s$ & -0.02438929995556706D0                    & 3.289201683319401D-05                & 0.005487672382653153D0  & -0.18967205251280128D0 & 1.3486331956279841D-05  & 0.037541628609521455D0 \\
        $\bar{u}s$ & -0.008308403859409175D0                   & 4.0866452739329755D-05               & 0.0026511859056011886D0 & -0.16896121633864306D0 & 2.2335584697813463D-05  & 0.033933534398936774D0 \\
        $\bar{c}d$ & -0.0620312387951807D0                     & 2.267517735452839D-05                & 0.012612883806140607D0  & -0.25812436133618843D0 & -4.261253080856257D-06  & 0.050327388108494284D0 \\
        $\bar{u}b$ & -0.023580025808217338D0                   & 3.0657222293821745D-05               & 0.005352044311661076D0  & -0.19271439731198028D0 & 8.241909640369068D-06   & 0.03809804952102698D0  \\
        $\bar{c}b$ & -0.012801225226009098D0                   & 3.236168398393879D-05                & 0.0032221044092733972D0 & -0.1848252094694455D0  & 6.595742525862169D-06   & 0.036528211472614765D0 \\
        \bottomrule
    \end{tabular}
    \caption{Fit coefficients of $\mathcal{A}_\text{initial}$ in the two trigger categories (left/right) for a neutral (top), positively charge (middle), and negatively charged (bottom) particle for the different initial state flavours}
    \label{tab:acccoeffPre}
\end{table}

\begin{table}
    \sisetup{
        detect-mode,
        round-mode              = figures,
        round-precision         = 3,
        group-digits=false
    }
    \centering
    \begin{tabular}{l@{\hspace{1cm}}S[scientific-notation=false,table-format=-1.4,round-mode=places,round-precision=4]S[scientific-notation=true,table-format=-1.2e-1]S[scientific-notation=true,table-format=-1.2e-1]@{\hspace{1cm}}S[table-format=1.3,round-mode=places]S[scientific-notation=true,table-format=1.2e-1]S[table-format=-1.4,round-mode=places,round-precision=4]}
        \toprule
                   & \multicolumn{3}{c}{single photon trigger} & \multicolumn{3}{c}{combined trigger}                                                                                                       \\
                   & {$a$}                                     & {$b$}                                & {$c$}                   & {$a$}                & {$b$}                    & {$c$}                   \\
        \midrule
        $uu$       & 0.38525532656239947D0                     & -2.3628125387441066D-05              & 0.036583038825438785D0  & 1.4179009959807805D0 & 0.00019318402078541907D0 & -0.13213062021217128D0  \\
        $dd$       & 0.39943561504923303D0                     & 1.784454676602849D-05                & 0.03677608469507186D0   & 1.5431432938623932D0 & 0.0002603304761813639D0  & -0.15281513143882092D0  \\
        $cc$       & -0.08698019570012273D0                    & -0.00017882404715921708D0            & 0.11167279264415338D0   & 1.2978470088122862D0 & 0.00015943440694614355D0 & -0.11976935313211168D0  \\
        $ss$       & 0.5248353167619432D0                      & 4.6061903293067066D-05               & 0.01282638879632233D0   & 1.4554725336733858D0 & 0.00022367903663272575D0 & -0.1380001570888785D0   \\
        $bb$       & 0.14312123365961413D0                     & -3.5483692688954893D-05              & 0.060523486229233724D0  & 1.4594279432344206D0 & 0.0002497796055038875D0  & -0.15908339475374172D0  \\
        $uc$       & 0.4373782503387766D0                      & -2.1099726560508866D-05              & 0.023776375433901743D0  & 1.352570526510758D0  & 0.00017897228740841698D0 & -0.12577105956572918D0  \\
        $ds$       & 0.09950722594177391D0                     & -5.9714373125369575D-05              & 0.08759342621226496D0   & 1.2528303938956713D0 & 0.0001800583268401668D0  & -0.10282440526626438D0  \\
        $db$       & 0.2548839475478191D0                      & 1.2711069457280673D-05               & 0.04845909024311625D0   & 1.3204031551147875D0 & 0.00022174772906246543D0 & -0.12586100686977048D0  \\
        $sb$       & 0.6817853117541732D0                      & 9.064290919677945D-05                & -0.023320207176590838D0 & 1.6547033810359737D0 & 0.00029549146128929273D0 & -0.1829824410338565D0   \\
        \midrule
        $u\bar{d}$ & 0.28920582367011693D0                     & -5.392197698683376D-05               & 0.059170447528441354D0  & 1.288081714061273D0  & 0.00016647838199530424D0 & -0.10550214628762097D0  \\
        $c\bar{s}$ & 0.1360543536099816D0                      & -8.8595728151652D-05                 & 0.07775522896899756D0   & 1.1585741495362076D0 & 0.0001499435859582795D0  & -0.09041290604805757D0  \\
        $u\bar{s}$ & 0.36403229644970786D0                     & -2.0313192545953587D-05              & 0.04273987985113689D0   & 1.2135143760634988D0 & 0.00016756223554864104D0 & -0.09590287762626909D0  \\
        $c\bar{d}$ & 0.03753114628032014D0                     & -0.00010725799982331615D0            & 0.09494959625977027D0   & 1.2909510263482682D0 & 0.00018627160475987738D0 & -0.11257857644665563D0  \\
        $u\bar{b}$ & -0.058835087810597805D0                   & -0.00014543013924738662D0            & 0.10940912532011073D0   & 1.2017501461720643D0 & 0.00016579167960714954D0 & -0.10232834425581878D0  \\
        $c\bar{b}$ & -0.12533421966904126D0                    & -0.00015829179090177497D0            & 0.11481519723740238D0   & 1.140050828111984D0  & 0.00013845098165656946D0 & -0.09551852523990538D0  \\
        \midrule
        $\bar{u}d$ & 0.5044416517236326D0                      & 3.185865787709505D-05                & 0.017651476316145345D0  & 1.3818977638832035D0 & 0.00020755445448888415D0 & -0.12381726801188014D0  \\
        $\bar{c}s$ & -0.058784400414788184D0                   & -0.00018154856031669657D0            & 0.11332390252462604D0   & 0.8143455081368138D0 & 2.4380225860519337D-05   & -0.02858203998980347D0  \\
        $\bar{u}s$ & 0.16119824265350222D0                     & -7.56869322478831D-05                & 0.07597218513158863D0   & 1.24000868172595D0   & 0.00015736056718772585D0 & -0.09943870123404218D0  \\
        $\bar{c}d$ & 0.710682341234595D0                       & 0.00011764645546976406D0             & -0.025532282301762386D0 & 1.039022396261481D0  & 0.0001343541213216677D0  & -0.06984861338158219D0  \\
        $\bar{u}b$ & 0.5598552848055856D0                      & 6.35468637968493D-05                 & -0.002042606314717672D0 & 1.6368519487358948D0 & 0.00029132603866931765D0 & -0.1791466833427859D0   \\
        $\bar{c}b$ & 0.1535119999958045D0                      & -7.130706112897657D-05               & 0.06527192110263923D0   & 0.9315940135101826D0 & 9.266106762835439D-05    & -0.059585540188055924D0 \\
        \bottomrule
    \end{tabular}
    \caption{Fit coefficients of $\mathcal{A}_{m_{jj}}$ in the two trigger categories (left/right) for a neutral (top), positively charge (middle), and negatively charged (bottom) particle for the different final state flavours}
    \label{tab:acccoeffPost}
\end{table}

The additional acceptances of the $b$-tagged category are given by
    {
        \sisetup{
            detect-mode,
            round-mode              = figures,
            round-precision         = 3,
            group-digits=false,
        }
        \begin{align}
            \mathcal{A}_{b\text{-tag}}^\text{single photon} & = \num{0.7349093272718935D0} - \num{3.918931699816906D-06}\frac{m-\SI{225}{\GeV}}{\si{\GeV}} + \num{0.028483738485628442D0}\log\frac{m_S}{\si{\GeV}} \\
            \mathcal{A}_{b\text{-tag}}^\text{combined}      & =\num{0.301508632706373D0} - \num{0.00011289893741976084D0}\frac{m-\SI{225}{\GeV}}{\si{\GeV}} + \num{0.10288525863298086D0}\log\frac{m_S}{\si{\GeV}}
        \end{align}
    }

The mass transformation itself is parametrized as
\begin{equation}
    m_S = a + b\; \left(m_S-\SI{225}{\GeV}\right) + c\; \log\frac{m_S}{\si{\GeV}}\label{eq:massfit}\,,
\end{equation}
where the corresponding coefficients are given in \cref{tab:masscoeffs}.

\begin{table}
    \sisetup{
        detect-mode,
        round-mode              = places,
        group-digits=false,
    }
    \centering
    \begin{tabular}{l@{\hspace{1cm}}S[round-precision=1,table-format=3.1]S[round-precision=3,table-format=1.3]S[round-precision=2,table-format=-2.2]@{\hspace{1cm}}S[round-precision=1,table-format=3.1]S[round-precision=3,table-format=1.3]S[round-precision=2,table-format=-2.2]}
        \toprule
             & \multicolumn{3}{c}{single photon trigger} & \multicolumn{3}{c}{combined trigger}                                                                                               \\
             & {$a$}                                     & {$b$}                                & {$c$}                 & {$a$}                & {$b$}                & {$c$}                 \\
             & \si{\GeV}                                 &                                      & \si{\GeV}             & \si{\GeV}            &                      & \si{\GeV}             \\
        \midrule
        $uu$ & 281.4319562313968D0                       & 0.9673914266961734D0                 & -12.235462255102286D0 & 268.6419095201171D0  & 0.9600671239644399D0 & -9.859514713821117D0  \\
        $dd$ & 258.81946927233946D0                      & 0.9633838763717543D0                 & -8.394837719212434D0  & 272.21432619456056D0 & 0.9643979603066067D0 & -10.626478039631243D0 \\
        $cc$ & 224.11771622447955D0                      & 0.9529383908273734D0                 & -2.1334395373697994D0 & 254.35644581068794D0 & 0.9611958334854936D0 & -7.505911098316428D0  \\
        $ss$ & 284.55872258550187D0                      & 0.9525467718329332D0                 & -12.822919404917098D0 & 254.71289010244473D0 & 0.9425188137399285D0 & -7.543214214890992D0  \\
        $bb$ & 220.9499460459883D0                       & 0.9266918231874117D0                 & -2.069497910863455D0  & 267.9237702016445D0  & 0.9388069130521747D0 & -10.263128783184035D0 \\
        $uc$ & 289.3880456860291D0                       & 0.9629350202691872D0                 & -13.78476867531609D0  & 280.28817594827103D0 & 0.9587600118446112D0 & -12.115839738882553D0 \\
        $ds$ & 193.21640722258093D0                      & 0.9349332966650078D0                 & 3.7982633636725636D0  & 223.71851983237093D0 & 0.9454616308413796D0 & -1.7582329141468864D0 \\
        $db$ & 216.08451117396797D0                      & 0.9327140780790565D0                 & -0.8275101427359572D0 & 233.5373585585733D0  & 0.9367802595332158D0 & -3.953300640830744D0  \\
        $sb$ & 238.35298494195436D0                      & 0.9419221069111677D0                 & -4.847633989956753D0  & 229.9860946965195D0  & 0.9387178657637238D0 & -3.373496593320827D0  \\
        \midrule
        $ud$ & 272.117378980746D0                        & 0.9685260914441841D0                 & -10.599655101495404D0 & 276.0391663421334D0  & 0.9661644772641924D0 & -11.14797232157632D0  \\
        $cs$ & 258.9945452730261D0                       & 0.9564116331738203D0                 & -8.177737850503428D0  & 242.3824874015853D0  & 0.9481249846138476D0 & -5.194289746546004D0  \\
        $us$ & 154.2868145807532D0                       & 0.923627818671771D0                  & 10.522652458614687D0  & 221.30860410941688D0 & 0.9471664169027666D0 & -1.4268813388883332D0 \\
        $cd$ & 243.18626334632324D0                      & 0.9531014690805794D0                 & -5.539394174464264D0  & 225.17451363750908D0 & 0.9431376149033346D0 & -2.241990843366313D0  \\
        $ub$ & 273.31450166305825D0                      & 0.954571643579162D0                  & -11.148508503812018D0 & 264.80736830157275D0 & 0.9503271038762466D0 & -9.478877550413694D0  \\
        $cb$ & 209.0255570171927D0                       & 0.9211762223473134D0                 & 0.6774235947175775D0  & 253.24791198178133D0 & 0.9382083805622718D0 & -7.390505995052736D0  \\
        \bottomrule
    \end{tabular}
    \caption{Fit coefficients of the mass transformation function \cref{eq:massfit} in the two trigger categories (left/right) for neutral (top) and charged (bottom) particles and the different final state flavours.}
    \label{tab:masscoeffs}
\end{table}


\section{Applying \texttt{HiggsBounds} to exotic scalars: A User's Guide}%
\label{sec:HB}

In this Appendix we provide a brief documentation of the new functions and subroutines that we added to \HiBo in \HiBov{5.10.0} to enable the application to neutral and charged scalars with an exotic, non-Higgs-boson-like coupling structure. We refer to \ccite{Bechtle:2020pkv} for the full documentation of the \HiBov{5} code.

\subsection{Extension of the Theoretical Input Framework}

We extended the standard \HiBo\ theoretical input framework both for the neutral and charged scalars. The cross sections for exotic production channels at the LHC at $\SI{13}{\TeV}$ are parametrized as fit functions that take effective couplings as input, as described above. For neutral scalars, we therefore introduce two new input subroutine for the neutral scalar couplings: First, the subroutine \Code{HiggsBounds_neutral_input_effC_firstgen}, which takes the couplings to first-generation fermions as input, which are defined as SM-Higgs-boson-normalized quantities, to be in accordance with the standard \HiBo\ effective coupling input:
\begin{align}
\mathcal{L}_{\text{Yukawa}}^{1^\text{st}~\text{gen.}} = \sum_{f = u, d, e} y_f \, \overline{f} (g_{\phi ff}^s + g_{\phi ff}^p \gamma^5  ) f \phi,
\end{align}
with $y_f = m_f/v$ and $v=\SI{174.1}{GeV}$. The up-quark, down-quark and electron masses are set to $m_u = \SI{2.16}{MeV}$, $m_d = \SI{4.67}{MeV}$ and $m_e = \SI{0.511}{MeV}$, respectively.

Second, the subroutine \Code{HiggsBounds_neutral_input_effC_FV} takes the input for the flavor-violating (FV) couplings to quarks.\footnote{Lepton-flavor violating decay modes of neutral scalars are already accounted for in the standard \HiBo input framework via the subroutine \Code{HiggsBounds_neutral_input_nonSMBR}~\cite{Bechtle:2020pkv}.} As these couplings vanish for the SM Higgs boson, they are not normalized and defined as $g,\tilde{g}$ from \cref{eq:neutral_scalar_model}.

In the standard \HiBo framework the branching ratios for neutral Higgs bosons can be approximated from the provided effective coupling through the rescaling of the corresponding partial width predictions from the LHC Higgs Cross Section Working Group (LHC HXSWG). As these predictions do neither exist for first generation fermionic final states, nor for flavor-violating fermionic final states, the corresponding branching ratios (BR) have to be set directly as input.
The input for the branching ratios of neutral scalar boson two-body decays to first generation fermions and FV quark final states can be provided via the new subroutines \Code{HiggsBounds_neutral_input_firstgenBR} and \Code{HiggsBounds_neutral_input_FVBR}, respectively, see \cref{tab:subroutines} for details.

For charged scalar bosons the effective couplings can be specified via the subroutine \Code{HiggsBounds_charged_input_effC_fermions}. These are defined as
$g_{qL}, g_{qR}$ from \cref{eq:charged_scalar_model}.
The branching ratios for charged scalar boson decays to fermionic final states $ud$, $us$, $cd$, $ub$, $e\nu$ and $\mu\nu$ can be set via the subroutine \Code{HiggsBounds_charged_input_firstgenBR}.

\begin{table}[t]
  \centering
\footnotesize
\begin{tabularx}{\textwidth}{ll}
  \toprule
  Subroutine name & Arguments \\
  \midrule
  \multirow{2}{*}{\Code{HiggsBounds_neutral_input_effC_firstgen}} &
  \dtypearray{dp}{$N$}~\Code{ghjuu_s}, \dtypearray{dp}{$N$}~\Code{ghjuu_p}, \dtypearray{dp}{$N$}~\Code{ghjdd_s},\\
  & \dtypearray{dp}{$N$}~\Code{ghjdd_p}, \dtypearray{dp}{$N$}~\Code{ghjee_s}, \dtypearray{dp}{$N$}~\Code{ghjee_p} \\[1em]
  \multirow{4}{*}{\Code{HiggsBounds_neutral_input_effC_FV}} &
  \dtypearray{dp}{$N$}~\Code{ghjuc_s}, \dtypearray{dp}{$N$}~\Code{ghjuc_p},\dtypearray{dp}{$N$}~\Code{ghjut_s}, \\
& \dtypearray{dp}{$N$}~\Code{ghjut_p}, \dtypearray{dp}{$N$}~\Code{ghjct_s}, \dtypearray{dp}{$N$}~\Code{ghjct_p},\\
& \dtypearray{dp}{$N$}~\Code{ghjds_s}, \dtypearray{dp}{$N$}~\Code{ghjds_p}, \dtypearray{dp}{$N$}~\Code{ghjdb_s},\\
& \dtypearray{dp}{$N$}~\Code{ghjdb_p},
 \dtypearray{dp}{$N$}~\Code{ghjsb_s},
 \dtypearray{dp}{$N$}~\Code{ghjsb_p}\\[1em]
\Code{HiggsBounds_neutral_input_firstgenBR} & \dtypearray{dp}{$N$}~\Code{BR_hjuu}, \dtypearray{dp}{$N$}~\Code{BR_hjdd}, \dtypearray{dp}{$N$}~\Code{BR_hjee} \\[1em]
\multirow{2}{*}{\Code{HiggsBounds_neutral_input_FVBR}} &
\dtypearray{dp}{$N$}~\Code{BR_hjuc}, \dtypearray{dp}{$N$}~\Code{BR_hjds}, \dtypearray{dp}{$N$}~\Code{BR_hjut},\\
& \dtypearray{dp}{$N$}~\Code{BR_hjdb}, \dtypearray{dp}{$N$}~\Code{BR_hjct}, \dtypearray{dp}{$N$}~\Code{BR_hjsb}\\[1em]
\multirow{6}{*}{\Code{HiggsBounds_charged_input_effC_fermions}} &
    \dtypearray{dp}{$M$}~\Code{hcjud_L}, \dtypearray{dp}{$M$}~\Code{hcjud_R}, \dtypearray{dp}{$M$}~\Code{hcjcs_L},\\
&    \dtypearray{dp}{$M$}~\Code{hcjcs_R}, \dtypearray{dp}{$M$}~\Code{hcjtb_L}, \dtypearray{dp}{$M$}~\Code{hcjtb_R},\\
&    \dtypearray{dp}{$M$}~\Code{hcjus_L}, \dtypearray{dp}{$M$}~\Code{hcjus_R}, \dtypearray{dp}{$M$}~\Code{hcjub_L},\\
&    \dtypearray{dp}{$M$}~\Code{hcjub_R}, \dtypearray{dp}{$M$}~\Code{hcjcd_L}, \dtypearray{dp}{$M$}~\Code{hcjcd_R},\\
&    \dtypearray{dp}{$M$}~\Code{hcjcb_L}, \dtypearray{dp}{$M$}~\Code{hcjcb_R}, \dtypearray{dp}{$M$}~\Code{hcjtd_L},\\
&    \dtypearray{dp}{$M$}~\Code{hcjtd_R}, \dtypearray{dp}{$M$}~\Code{hcjts_L}, \dtypearray{dp}{$M$}~\Code{hcjts_R}\\[1em]
\multirow{2}{*}{\Code{HiggsBounds_charged_input_firstgenBR}} & \dtypearray{dp}{$M$}~\Code{BR_Hpjud}, \dtypearray{dp}{$M$}~\Code{BR_Hpjus}, \dtypearray{dp}{$M$}~\Code{BR_Hpjcd},\\
& \dtypearray{dp}{$M$}~\Code{BR_Hpjub}, \dtypearray{dp}{$M$}~\Code{BR_Hpjenu}, \dtypearray{dp}{$M$}~\Code{BR_Hpjmunu}\\
  \bottomrule
\end{tabularx}
\caption{Listing of all new \HiBo\ input subroutines.  The subroutine arguments (right column) are specified by their default name and their Fortran data type (\emph{gray font}) as double precision (\texttt{dp}) with array length quoted in parentheses. The number of neutral and charged scalar bosons in the model are denoted by $N$ and $M$, respectively.}%
\label{tab:subroutines}
\end{table}


\clearpage
\printbibliography

\end{document}